\documentclass[aps,pra,epsf,amsmath,superscriptaddress,citeautoscript,showpacs,floatfix,twocolumn,nofootinbib]{revtex4-2}
\usepackage[T1]{fontenc}
\usepackage{txfonts}

\usepackage[dvipsnames]{xcolor}
\definecolor{red}{rgb}{0.9, 0,0}
\definecolor{cerulean}{rgb}{0., 0.42,0.9}
\definecolor{navy}{rgb}{0.05, 0.05,0.8}

\usepackage{setspace}
\usepackage[colorlinks]{hyperref}
\hypersetup{
    colorlinks = true,
    citecolor  = red,
	linkcolor  = navy
}

\usepackage{slashed}
\usepackage{graphics}
\usepackage{amsmath}
\usepackage{amssymb}
\usepackage{latexsym}
\usepackage{mathtools}
\usepackage{dsfont}
\usepackage{cleveref}
\usepackage{amsfonts}
\usepackage{enumitem}
\usepackage{xstring} 
\usepackage{xspace} 
\usepackage[normalem]{ulem}
\usepackage{fontawesome}
\usepackage{braket}
\usepackage{mathrsfs}
\usepackage{bm}
\usepackage{comment}

\newcommand{\be}{\begin{equation}}
\newcommand{\ee}{\end{equation}}
\newcommand{\bs}{\begin{split}}
\newcommand{\es}{\end{split}}

\usepackage{subfiles} 

\begin{document}

\title{Amplification and generation bounds of gravity-induced entanglement in pulsed optomechanical systems}
\author{Daisuke Miki}
\email{dmiki@caltech.edu}
\affiliation{Burke Institute for Theoretical Physics, California Institute of Technology, Pasadena, California 91125, USA}
\author{Alfred Li}
\email{ali2@caltech.edu}
\affiliation{Burke Institute for Theoretical Physics, California Institute of Technology, Pasadena, California 91125, USA}
\author{Yanbei Chen}
\email{yanbei@caltech.edu}
\affiliation{Burke Institute for Theoretical Physics, California Institute of Technology, Pasadena, California 91125, USA}

\date{\today}

\begin{abstract}
  {  We investigate gravity-induced entanglement between the output optical fields of two red-detuned pulsed optomechanical systems with their masses coupled by mutual gravitational interaction. For each individual system, the optomechanical interaction realizes a beam-splitter state swap between an incident optical pulse and its mechanical mode. Using two rectangular pulses for each system ---the first to imprint a nonclassical state on the mechanical modes and the second to read the gravitationally generated entanglement back onto the outgoing light---we show that the amount of  entanglement can be amplified by preparing the input in a squeezed or Fock state. However, the threshold for entanglement generation is set by the competition between the gravitational coupling and thermal decoherence, $g_G>2\gamma_m N_{\rm th}$, and cannot be lowered by any choice of input state. We prove this bound for two-mode Gaussian inputs and show that it continues to hold for Fock-state inputs. We further analyze how imperfect detection modifies the threshold and identify the entanglement-annihilating and entanglement-breaking regimes, which are set by the thermal decoherence accumulated over the interaction time, independent of the gravitational coupling.}
\end{abstract}

\maketitle

%

\singlespacing 
\section{Introduction}

\label{sec:intro}
Formulating a quantum theory of gravity remains one of the central challenges in fundamental physics. Despite the remarkable success of both general relativity and quantum mechanics in their respective domains, direct experimental access to quantum gravity is widely believed to require energies near the Planck scale. Recently, however, it has been proposed that certain low-energy signatures of linearized quantum gravity may be accessible in the Newtonian regime through gravity-induced entanglement between two massive objects \cite{bose2017,marletto2017}. Since such entanglement cannot be generated by a classical gravitational interaction \cite{marletto2020,marletto2025,marletto2025-2}, these proposals have stimulated extensive discussion on the connection between gravity-induced entanglement and the quantization of the dynamical degrees of freedom of the gravitational field \cite{mari2016,belenchia2018,danielson2022,carney2022,sugiyama2023,sugiyama2024}. Proposed platforms for testing this idea include two-level systems \cite{bose2017,marletto2017,marshman2020,schut2024}, harmonic oscillators \cite{krisnanda2020,qvarfort2020}, and hybrid systems \cite{carney2021,matsumura2022}.

Among these, optomechanical systems have attracted particular interest \cite{balushi2018,matsumura2020,miki2022,miao2020,datta2021,miki2024,miki20242,mari2025}. Such systems couple macroscopic mechanical oscillators to optical fields and offer a promising route to both the generation and readout of gravity-induced entanglement. In suitably designed setups, gravitational interactions can become comparable to or even dominate over other 
interactions, such as the Casimir-Polder force. 
A major obstacle, however, is the extreme weakness of gravity. In measurement-based optomechanical schemes, explicit conditions for the generation of gravity-induced entanglement have been derived \cite{miao2020,datta2021,miki2024,miki20242}, which require the gravitational interaction to overcome thermal decoherence. Several directions have been explored to address or circumvent this difficulty, including optimization of noise sources and experimental configurations \cite{tang2025,matsumoto2025}, amplification of gravitational signatures \cite{pedernales2022,kaku2023,fujita2025,shiomatsu2025,hatakeyama2026,fukuzumi2026}, and schemes designed to distinguish quantum gravity from a semiclassical gravity model \cite{liu2023,liu2025,miki2025,zhong2025}.
Related approaches have also proposed tests that do not rely directly on the generation of entanglement, such as testing whether the gravitationally induced dynamics can be simulated by local operations and classical communication \cite{lami2024}, or whether gravity can mediate a nonclassical quantum information channel \cite{mari2025}.

In this paper, we propose a method to amplify gravity-induced entanglement using pulsed optomechanics, which provides
powerful tools for quantum-state preparation and tomography of mechanical motion \cite{vanner2011,hofer2011,bassam2019}. For a red-detuned cavity, the optomechanical interaction reduces to a beam-splitter-type Hamiltonian, which enables state transfer between an optical pulse and a mechanical mode. This makes it possible to prepare highly nonclassical mechanical states by appropriately engineering the input optical pulse. Ref.~\cite{jordan2024} studied semiclassical-gravity effects in a single pulsed optomechanical system. By contrast, here we focus on quantum effects of gravity and show that the amount of gravity-induced entanglement can be enhanced by injecting squeezed states or Fock states {of light}. Furthermore, the mechanically generated entanglement can be transferred back to a second optical pulse, enabling direct optical readout as entanglement between the two outgoing optical fields.

At the same time, our analysis reveals a fundamental limitation of such amplification strategies. Although nonclassical initial states can enhance the amount of entanglement once the system is already in the entangling regime, they do not shift the optimal boundary between entangling and non-entangling regimes. This threshold is determined solely by the competition between {the strength of} gravitational interaction and {the level of} thermal decoherence.
{In particular, previous continuous-measurement optomechanical schemes using coherent light found that gravity-induced entanglement requires $g_G>2\gamma_mN_{\rm th}$, where $g_G=GM/d^3\omega_m$, with
$\omega_m$ the bare mechanical resonance frequency, $\gamma_m$ the mechanical damping rate, defined as the half width at half maximum, and $N_{\rm th}\simeq k_B T/\hbar\omega_m$ the thermal occupation number \cite{miao2020,datta2021}.
\footnote{Refs.~\cite{miao2020,datta2021} also showed that gravity-induced squeezing of the output light can reach $10\log_{10}\!\left(1+GM/d^3\gamma_m^2\right)\,\mathrm{dB}$ on the timescale $t=\gamma_m^2d^3/2GM$, provided that the optomechanical coupling is sufficiently stronger than thermal decoherence. The thermal-noise requirement for observing this squeezing is less stringent than that for satisfying the entanglement condition above. However, it has also been shown that such gravity-induced squeezing can arise in classical-gravity models \cite{liu2023}; distinguishing the quantum and classical cases therefore requires more carefully designed measurement protocols \cite{liu2025,miki2025,zhong2025}.}
This inequality reduces to $\hbar GM/d^3>2\gamma_mk_BT$ and expresses the requirement that coherent gravitational coupling must dominate thermal decoherence. For two-mode Gaussian states, Refs.~\cite{kafri2013,kafri2014} showed that entanglement cannot be generated beyond this threshold. In the present work, we show that the same threshold governs our pulsed protocol: while suitable input states can saturate the bound and substantially enhance the amount of entanglement, they cannot lower the threshold itself. We also demonstrate that the same conclusion remains valid even for highly non-Gaussian initial states, such as mechanical Fock states. More general extensions to multimode Gaussian and non-Gaussian settings will be presented in Ref.~\cite{alfred2026}.  More broadly, such threshold effects were also predicted for entanglement between mechanical oscillators and outgoing light fields~\cite{direkci2024macroscopic,direkci2025characterizing,direkci2025universality}.
}

Furthermore, we derive two additional entanglement-related conditions for two-mode Gaussian states.
First, we derive the entanglement-annihilating and entanglement-breaking conditions \cite{horodecki2003,holevo2008,moravcikova2010,filippov2014}. The entanglement-annihilating condition characterizes when the channel maps any input state of the two systems to a separable output state, while the entanglement-breaking condition guarantees separability even when the system is initially entangled with an external ancilla. We show that these regimes arise once thermal noise injects $\mathcal{O}(1)$ phonons during the interaction time, and that these conditions are independent of the gravitational coupling.
Second, we analyze the separability-preservation condition in the presence of measurement loss. Imperfect detection of outgoing optical fields modifies the bound for the entanglement generation, yielding a bound for detectable entanglement. In this case, infinitely large input optical squeezing is no longer optimal for approaching the modified threshold, even though larger squeezing can still enhance the amount of entanglement in part of the parameter space. Instead, the optimal squeezing becomes loss-dependent, and the modified bound can be saturated by choosing the input squeezing appropriately.

This paper is organized as follows. In Sec.~\ref{sec:formulas}, we introduce the dynamics of a pulsed optomechanical system with a single mechanical oscillator. In Sec.~\ref{sec:twoopto}, we formulate a two-system protocol for gravity-induced entanglement and study the dependence of entanglement negativity on system parameters. In particular, we demonstrate that the amount of entanglement can be amplified by using nonclassical input states, and that there exists a bound for entanglement generation. 
In Sec.~\ref{sec:entanglement bound}, we discuss the universal condition for separability preservation, together with the entanglement-annihilating and entanglement-breaking conditions.
In Sec.~\ref{sec:loss}, we analyze the effects of measurement loss. In Sec.~\ref{sec:summary}, we summarize our conclusions.

\section{Single optomechanical system}
\label{sec:formulas}
In this section, we review the pulsed optomechanical system with a single mechanical mirror \cite{vanner2011,hofer2011}. The system consists of a mechanical oscillator with mass $M$ and resonant frequency $\omega_m$, acting as a movable mirror of an optical cavity with length $\ell$ and resonant frequency $\omega_c$, which in turn is driven by a carrier laser with frequency $\omega_0$.  The Hamiltonian of this system is given by
\begin{align}
    \hat H& =
    \frac{\hat p^2}{2M}+\frac{1}{2}M\omega_m^2\hat x^2
    +\hbar\omega_c\hat a^\dagger\hat a
    -\frac{\hbar\omega_c}{\ell}\hat x\hat a^\dagger\hat a\nonumber\\
    &\quad
    +i\hbar E(t)\left(e^{-i\omega_0t}\hat a^\dagger-e^{i\omega_0t}\hat a\right),
\end{align}
where $\hat x$ and $\hat p$ are the position and momentum operators of the mirror, satisfying $[\hat x,\hat p]=i\hbar$; 
$\hat a$ and $\hat a^\dagger$ are the creation and annihilation operators of the cavity's optical mode, satisfying $[\hat a,\hat a^\dagger]=1$.
The last term describes the driving laser, and $E(t)$ is a time-dependent driving strength.

To simplify the interaction Hamiltonian, we introduce the phonon operators as 
\begin{equation}
    \hat b=\sqrt{\frac{M\omega_m}{2\hbar}}\hat x+\frac{i\hat p}{\sqrt{2\hbar M\omega_m}},\; 
    \hat b^\dagger=\sqrt{\frac{M\omega_m}{2\hbar}}\hat x-\frac{i\hat p}{\sqrt{2\hbar M\omega_m}}
\end{equation}
satisfying $[\hat b,\hat b^\dagger]=1$.
We then break the photon annihilation operator into a time-dependent classical amplitude $\bar a(t)$ and a fluctuating piece $\hat a$, writing $\hat a\rightarrow \bar a+\hat a$, and only retain terms up to quadratic in $(a,a^\dagger,b,b^\dagger)$ in the Hamiltonian.
In the rotating frame at the laser frequency $\omega_0$, the Hamiltonian in the interaction picture then becomes
\begin{align}
\label{eqHI}
    \hat H_I&=
    -\hbar g(t)(e^{-i\omega_mt}\hat b+e^{i\omega_mt}\hat b^\dagger)(e^{i\Delta t}\hat a+e^{-i\Delta t}\hat a^\dagger).
\end{align}
where $\Delta=\omega_0-\omega_c$ is the cavity detuning and $g(t)=\omega_cx_{\rm zp}\bar a(t)/\ell$ is the time-dependent optomechanical coupling strength depending on the pulse protocol with the zero point fluctuation $x_{\rm zp}=\sqrt{\hbar/2M\omega_m}$. Here we only consider real-valued $\bar a(t)$, corresponding to an externally driven amplitude modulation, and have ignored constant terms in the Hamiltonian as well as driving terms that lead to $\bar a(t)$.

When the cavity is {\it red detuned} from the laser, with $\Delta =-\omega_m$,
the interaction can realize a beam-splitter-type state-swap between the optical and mechanical modes as the coupling strength $g(t)$ is turned on for a duration $t_P$. This can be done cleanly if we require  separations between three time scales: the pulse duration $t_P$ should be much longer than the cavity storage time $1/\kappa$, which in turn must be much longer than the oscillation time scale of the mirror $1/\omega_m$.  Given this separation, we can drop terms that oscillate at $\sim 2\omega_m$ in $\hat H_I$ [Eq.~\eqref{eqHI}] and obtain
\begin{align}
\label{eq:pulsedquadratures}
    \dot{\hat{b}}&=ig(t)\hat{a},\;
    \dot{\hat{a}}=ig(t)\hat{b}-\kappa\hat a+\sqrt{2\kappa}\hat a_{\rm in},\;\hat a_{\rm out}= \sqrt{2\kappa}\hat a-\hat a_{\rm in}
\end{align}
Note here that we have additionally applied the input-output formalism (see e.g., Ref.~\cite{yanbei2013}) where $\kappa$ is the decay rate of the cavity to the outgoing continuum, while $\hat a_{\rm in}$ and $\hat a_{\rm out}$ are the input and output field operators, with $[\hat a_{\rm in}(t),\hat a_{\rm in}^{\dagger}(t^\prime)]=[\hat a_{\rm out}(t),\hat a_{\rm out}^{\dagger}(t^\prime)]=\delta(t-t')$. In Eq.~\eqref{eq:pulsedquadratures},  we have ignored thermal force noise acting on the mirrors.
In later sections, we shall account for thermal noise during the gravitational interaction stage between the two mirrors, while we do not explicitly treat thermal noise during the optomechanical state-transfer stage; since gravity is much weaker than the optomechanical interaction, and the relevant evolution time is much longer, thermal noise from that stage dominates.  Instead, we will include detection loss in our treatment in Sec.~\ref{sec:loss}.

To study state transfer, let us consider a single rectangular pulse, with the coupling strength given by $g(t)=(2n+1)(\pi/2t_P){\rm rect}(t/t_P-1/2)$, where the positive integer $n$ fixes the pulse area. This choice yields a constant coupling $g(t)=g$ during $0\le t<t_P$ and $g(t)=0$ otherwise. In this case, we can solve the equation during the pulse interaction in the limit $\kappa\gg g$ as
\begin{align}
    &\hat a=i\frac{g}{\kappa}\hat b+\sqrt{\frac{2}{\kappa}}\hat a_{\rm in}\\
    &\hat b(t)=e^{-Gt}\hat b(0)+i\sqrt{2G}\int_0^t ds\,e^{-G(t-s)}\hat a_{\rm in}(s),
\end{align}
where $G=g^2/\kappa$.
The output optical field is obtained by the input-output relation $\hat a_{\rm out}=\sqrt{2\kappa}\hat a-\hat a_{\rm in}$. We define the normalized temporal mode as
\begin{align}
    \hat \xi_{\rm in}&=
    \sqrt{\frac{2G}{e^{2Gt_P}-1}}\int_0^{t_P}dse^{Gs}\hat a_{\rm in}(s)\\
    \hat \xi_{\rm out}&=
    \sqrt{\frac{2G}{1-e^{-2Gt_P}}}\int_0^{t_P}dse^{-Gs}\hat a_{\rm out}(s),
\end{align}
where $[\hat \xi_{\rm in},\hat \xi_{\rm in}^\dagger]=[\hat \xi_{\rm out},\hat \xi_{\rm out}^\dagger]=1$. Defining the initial phonon operator as $\hat b(0)=\hat b_{\rm in}$, we have
\begin{align}
    \hat b(t_P)&=e^{-Gt_P}\hat b_{\rm in}+i\sqrt{1-e^{-2Gt_P}}\hat \xi_{\rm in}\\
    \hat \xi_{\rm out}&=i\sqrt{1-e^{-2Gt_P}}\hat b_{\rm in}+e^{-Gt_P}\hat \xi_{\rm in}.
\end{align}
In the strong coupling regime of $Gt_P>1$, we can approximate $e^{-Gt_P}\simeq0$, which leads to $\hat b(t_P)\simeq i\hat \xi_{\rm in}$ and $\hat \xi_{\rm out}\simeq i\hat b_{\rm in}$. This corresponds to a complete state swap between the mechanical mode and the optical mode. For instance, when the input $n$-photon state and the mechanical ground state is given by $\ket{\Psi_{\rm in}}=\ket{0}_m\ket{n}_p$, the output state is $\ket{\Psi_{\rm out}}=\ket{n}_m\ket{0}_p$. In the present paper, we only focus on the complete state swap case for simplicity.

\begin{figure}[b]
    \centering
    \includegraphics[width=8.0cm]{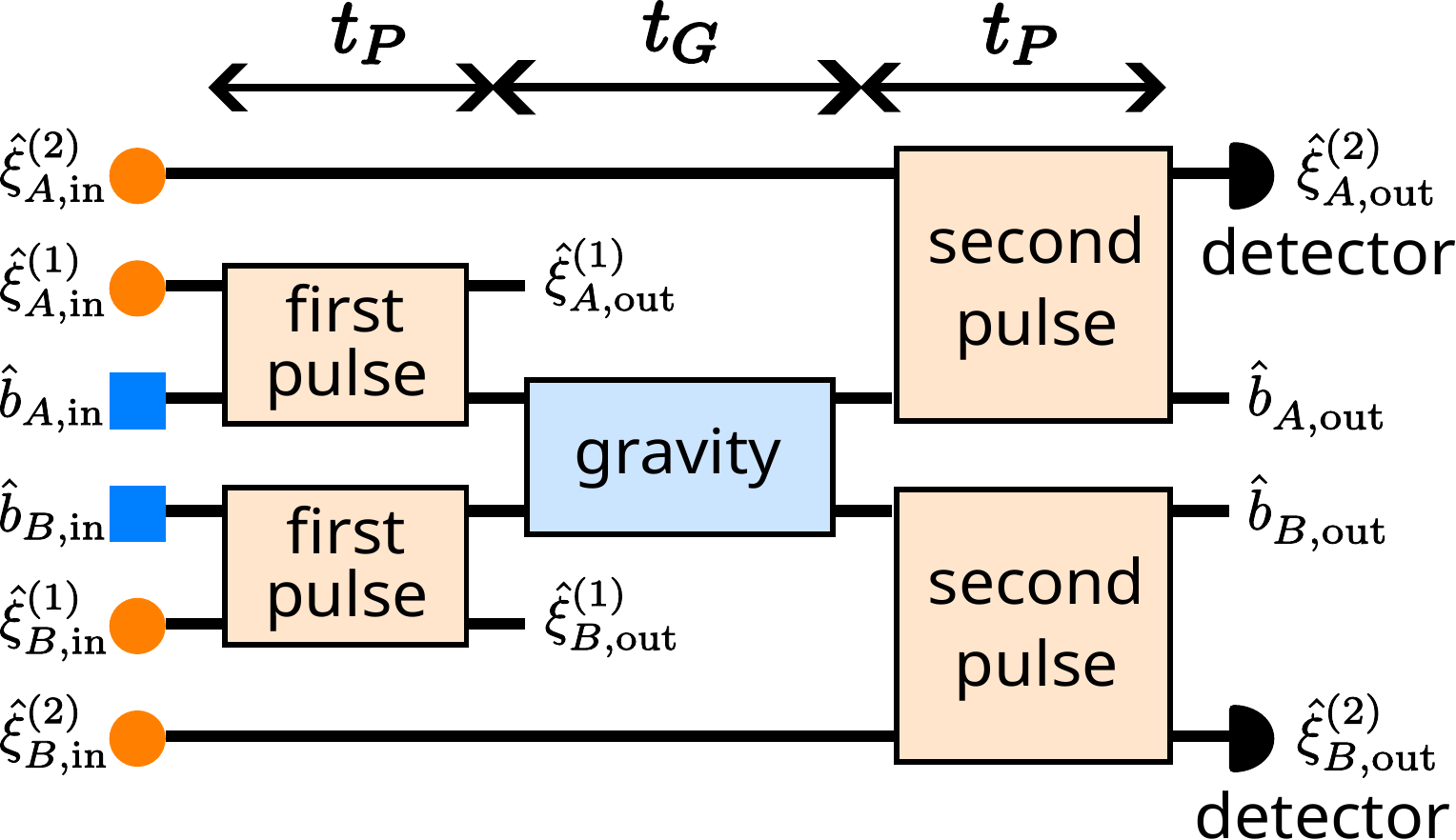}
    \caption{Schematic diagram of our protocol, which consists of three stages, with time going to the right: first pulse (duration $t_P$), gravitational interaction (duration $t_G$), and second pulse (duration $t_P$). From top left to bottom left, we have six input operators, for the second and first pulses of mass A, mass A, mass B, and first and second pulses of mass B. At the right, the top most and the bottom most operators are outgoing optical fields from second pulses of masses A and B, respectively.  They are the only fields detected.}
    \label{fig:pulse}
\end{figure}

\section{Gravity-induced entanglement in two optomechanical systems}
\label{sec:twoopto}
One advantage of the pulsed optomechanical platform is that it enables the preparation of highly nonclassical mechanical states by swapping suitably engineered optical input states onto the mirrors, as well as the characterization of gravity-induced entanglement via optical detection.
Depending on the choice of initial states, this strategy
can lead to a more efficient confirmation of gravity-induced entanglement.

In this section, after introducing our protocol and its basic equations, we consider gravity-induced entanglement of outgoing light fields when input light fields are in Gaussian and Fock states, respectively.

\subsection{Protocol and Input-Output Relation}
Our protocol consists of three steps, with a pulse sequence illustrated in  Fig.~\ref{fig:pulse}.  During the first stage, mass A and mass B each separately interact with their respective input optical field via pulsed optomechanics. The {\it first pulse} is designed to swap the state of the input optical fields onto the masses. {\it The input fields A and B are independent (separable).} Next, the two mechanical systems {\it gravitationally interact} for a duration  of $t_G$, with the intention of creating quantum entanglement.  Finally, a {\it second pulse} on each of the masses swaps the masses' quantum states onto their respective outgoing optical fields for readout.  Any entanglement between the outgoing fields A and B would have been induced by gravity. We ignore both thermal noise and gravitational interaction during the pulsed optomechanical stages, as they are designed to be short compared with the gravitational interaction stage.
Although this pulse sequence can enhance the amount of entanglement, we show that environmental thermal noise during the gravitational interaction severely limits the generation of gravity-induced entanglement.

We assume the quantum gravitational interaction in the Newtonian limit to have a Hamiltonian of
\begin{align}
    \hat H_G=-\frac{GM_AM_B}{d-\hat x_A+\hat x_B},
\end{align}
where $G$ is the gravitational constant and $d$ is the separation between two mirrors. This interaction arises naturally in linearized quantum gravity in the nonrelativistic Newtonian limit. Assuming the two oscillators to have the same $\omega_m$, and assuming an interaction time much longer than the period of oscillation, then, in the interaction picture and rotating-wave approximation, we can write an interaction Hamiltonian of
\begin{align}
    \hat H_{G,I}=\hbar g_G(\hat b_A\hat b_B^\dagger+\hat b_A^\dagger\hat b_B),
    \label{gravint}
\end{align}
where $g_G=GM/d^3\omega_m$ is the strength of the gravitational interaction for the symmetric setup.
The other terms of the Hamiltonian are the same as the previous section (also see Appendix~\ref{apdx:Hamiltonian}).

For the two state-transfer stages (labeled ``first pulse'' and ``second pulse'' in Fig.~\ref{fig:pulse}) we consider the two rectangular pulses, whose coupling strengths are given by
\begin{align}
    g_j(t)&=
    \pi\frac{2n_j+1}{2t_P}
    \left[
    {\rm rect}\left(\frac{t-t_P/2}{t_P}\right)\right.\nonumber\\
    &\left.+{\rm rect}\left(\frac{t-t_P/2-t_G-t_P}{t_P}\right)
    \right],
\end{align}
with a positive integer $n_j$. During the first pulse $0\le t<t_P$ and the second pulse $t_P+t_G\le t<2t_P+t_G$, we ignore the gravitational interaction and mechanical thermal noise, since the optomechanical interaction is assumed to be much stronger than both the gravitational interaction and the effect of mechanical thermal noise during these intervals.

We focus here on the gravitational-interaction stage (labeled as ``gravity'' in Fig.~\ref{fig:pulse}), during which the two mechanical oscillators evolve under their mutual gravitational interaction. Since gravity is extremely weak, generating entanglement requires a relatively long interaction time, and thermal noise must then be taken into account. The two mechanical oscillators therefore evolve as
\begin{align}
    &\frac{d}{dt}\hat b_A=
    -ig_G\hat b_B-\gamma_m\hat b_A+f_{A,{\rm th}}\\
    &\frac{d}{dt}\hat b_B=
    -ig_G\hat b_A-\gamma_m\hat b_B+f_{B,{\rm th}},
\end{align}
where $\gamma_m$ is the mechanical decay rate, defined as the half width at half maximum, and $f_{j,{\rm th}}$ is the thermal fluctuation of mass $j$, with zero mean and a two-time correlation function of
$\braket{\{f_{j,{\rm th}}(t),f_{j,{\rm th}}(t^\prime)^\dagger\}}/2\simeq2\gamma_mN_{\rm th}\delta(t-t^\prime)$, where $N_{\rm th}\simeq k_BT/\hbar\omega_m$ is the equilibrium thermal occupation number of each free oscillator at  the high temperature limit. Then, the output operators are
\begin{align}
    &\hat b_A(t_P+t_G)=
    C_{11}(t_G)\hat b_A(t_P)+C_{12}(t_G)\hat b_B(t_P)+n_A\\
    &\hat b_B(t_P+t_G)=
    C_{21}(t_G)\hat b_A(t_P)+C_{22}(t_G)\hat b_B(t_P)+n_B,
\end{align}
where the coefficients $C_{ij}(t)$ are given by
\begin{align}
\begin{split}
    &C_{11}(t)=C_{22}(t)=e^{-\gamma_mt}\cos[g_Gt]\\
    &C_{12}(t)=C_{21}(t)=-ie^{-\gamma_mt}\sin[g_Gt]
    \label{cmat}
\end{split}
\end{align}
\begin{align}
    n_A&=
    \int_{t_P}^{t_P+t_G}ds(C_{11}(t_P+t_G-s)f_{A,{\rm th}}(s)\nonumber\\
    &\qquad\qquad\qquad+C_{12}(t_P+t_G-s)f_{B,{\rm th}}(s))\\
    n_B&=
    \int_{t_P}^{t_P+t_G}ds(C_{21}(t_P+t_G-s)f_{A,{\rm th}}(s)\nonumber\\
    &\qquad\qquad\qquad+C_{22}(t_P+t_G-s)f_{B,{\rm th}}(s)).
\end{align}
We finally derive the output operators of the second optical modes as (also see Appendix~\ref{apdx:Hamiltonian})
\begin{align}
    \label{xiathermal}
    &\hat \xi_{A,{\rm out}}^{(2)}=
    -C_{11}(t_G)\hat \xi_{A,{\rm in}}^{(1)}+C_{12}(t_G)\hat \xi_{B,{\rm in}}^{(1)}+in_A\\
    &\hat \xi_{B,{\rm out}}^{(2)}=
    C_{21}(t_G)\hat \xi_{A,{\rm in}}^{(1)}-C_{22}(t_G)\hat \xi_{B,{\rm in}}^{(1)}-in_B.
    \label{xibthermal}
\end{align}

\subsection{Initial Gaussian states}
\label{sec:initial gaussian states}
Having obtained the input-output relation of our protocol, we shall consider the output state achievable from our protocol when different input states are used.  Let us first consider injecting Gaussian input optical states during the first pulse stage. Linearity of the dynamics produces a Gaussian state for the outgoing fields, which is fully characterized by their covariance matrix.
To characterize output entanglement, let us introduce the amplitude quadrature $\hat X_{j,{\rm in}/{\rm out}}^{(1/2)}=\hat \xi_{j,{\rm in}/{\rm out}}^{(1/2)}+\hat \xi_{j,{\rm in}/{\rm out}}^{(1/2)\dagger}$ and phase quadrature $\hat Y_{j,{\rm in}/{\rm out}}^{(1/2)}=(\hat \xi_{j,{\rm in}/{\rm out}}^{(1/2)}-\hat \xi_{j,{\rm in}/{\rm out}}^{(1/2)\dagger})/i$ satisfying $[\hat X_{j,{\rm in}/{\rm out}}^{(1/2)},\hat Y_{j,{\rm in}/{\rm out}}^{(1/2)}]=2i$. Defining vectors $\hat{\bm R}_{{\rm in}/{\rm out}}^{(1/2)}=(\hat X_{A,{\rm in}/{\rm out}}^{(1/2)},\hat Y_{A,{\rm in}/{\rm out}}^{(1/2)},\hat X_{B,{\rm in}/{\rm out}}^{(1/2)},\hat Y_{B,{\rm in}/{\rm out}}^{(1/2)})^{\rm T}$, the equation can be written as $\hat{\bm R}_{\rm out}^{(2)}=\bm M\hat{\bm R}_{\rm in}^{(1)}+\bm n_{\rm th}$, where
\begin{align}
    \bm M=\left(\begin{array}{cccc}
    -C_{11}(t_G) & 0 & 0 & iC_{12}(t_G) \\
    0 & -C_{11}(t_G) & -iC_{12}(t_G) & 0 \\
    0 & iC_{21}(t_G) & -C_{22}(t_G) & 0 \\
    -iC_{21}(t_G) & 0 & 0 & -C_{22}(t_G)
    \end{array}\right),
\end{align}
and $\bm n=(n_{X_A},n_{Y_A},n_{X_B},n_{Y_B})^{\rm T}$ with $n_{X_j}=i(n_j-n_j^{\dagger})$ and $n_{Y_j}=n_j+n_j^{\dagger}$.
Then, the covariance matrix of the output optical fields $\bm V_{\rm out}=\braket{\{\hat{\bm R}_{\rm out}-\braket{\hat{\bm R}_{\rm out}},(\hat{\bm R}_{\rm out}-\braket{\hat{\bm R}_{\rm out}})^{\rm T}\}}/2$ is
\begin{align}
\label{double squeezed CM}
    \bm{V}_{\rm out}&=
    \bm M\bm V_{\rm in}\bm M^{\rm T}+\bm V_{\rm th},\quad
    \bm V_{\rm th}=2
    (1-e^{-2\gamma_mt_G})N_{\rm th}\mathbf 1,
\end{align}
where $\bm V_{\rm in}$ is the initial covariance matrix of the first optical modes and $\bm V_{\rm th}$ represents the thermal noise effects. We introduce the entanglement negativity for the two-mode Gaussian states as $\mathcal N_G=(1/\nu_--1)/2$, where $\nu_-$ is the smallest symplectic eigenvalue defined by $\nu_-^2=(\sigma-\sqrt{\sigma^2-4{\rm det}\bm V_{\rm out}})/2$ with $\sigma=\det\bm V_{A,{\rm out}}+\det\bm V_{B,{\rm out}}-2\det\bm V_{AB,{\rm out}}$ and $\bm V_{\rm out}$ is written in terms of the local covariance blocks $\bm V_{A,{\rm out}}$ and $\bm V_{B,{\rm out}}$ and the correlation block $\bm V_{AB,{\rm out}}$. As we only have two single modes, entanglement exists between A and B  if and only if $\mathcal N_G>0$.

The gravitational interaction in Eq.~\eqref{gravint} has a beam-splitter form up to a local phase rotation of one of the modes by $\pi/2$. With our quadrature convention, the corresponding optimal initial condition is to have the two oscillators squeezed along the same direction, for example with $\bm V_{\rm in}={\rm diag}\{e^{2\zeta},e^{-2\zeta},e^{2\zeta},e^{-2\zeta}\}$.
This choice is only unique up to a common phase-space rotation: applying the same local rotation to both input squeezed states only produces the corresponding local rotation at the output and therefore does not change the amount of entanglement; what matters is the relative orientation of the two squeezing axes, which must be aligned in order to be optimal. Note that vacuum inputs remain separable, since a passive beam-splitter transformation cannot entangle two vacuum modes.

For this class of input, in the regime $\gamma_mt_G\ll1$, we obtain the entanglement negativity as
\begin{align}
    \mathcal N_G^{\rm sq}
    &\simeq
    \frac{1}{2}\biggl(
    \sqrt{
    \left(\cosh 2\zeta+4\gamma_mt_GN_{\rm th}\right)^2
    -\sinh^2 2\zeta\,\cos^2(2g_Gt_G)}\nonumber\\
    &\qquad-\left|\sinh 2\zeta\,\sin(2g_Gt_G)\right|\biggr)^{-1}
    -\frac{1}{2}.
    \label{ngsq_full}
\end{align}
The condition $\mathcal N_G^{\rm sq}>0$ becomes
\begin{align}
    4\gamma_mt_GN_{\rm th}<\sqrt{\cosh^2[2\zeta]+2|\sinh[2\zeta]\sin[2g_Gt_G]|}-\cosh[2\zeta].
    \label{ngsq_condition}
\end{align}
The right-hand side is a monotonically increasing function of
$|\zeta|$ and approaches $|\sin(2g_Gt_G)|$ in the limit of large squeezing. Therefore, for a given gravitational coupling, the maximum tolerable thermal noise within this squeezed-input family is obtained as $|\zeta|\to\infty$, where Eq.~\eqref{ngsq_condition} reduces to $4\gamma_mt_GN_{\rm th}<|\sin[2g_Gt_G]|$. Since $|\sin(2g_Gt_G)|\le 2g_Gt_G$ for $g_Gt_G\ge0$, entanglement generation implies
\begin{align}
    g_G>2\gamma_mN_{\rm th}.
    \label{gie}
\end{align}
Thus, Eq.~\eqref{gie} is a necessary condition that is independent of the initial state: if $g_G\le2\gamma_mN_{\rm th}$, no entanglement can be generated by this squeezed-input protocol for any squeezing strength. Conversely, for finite squeezing, Eq.~\eqref{gie} is not by itself sufficient; the actual condition is the stronger finite-squeezing condition \eqref{ngsq_condition}. Squeezing can increase the amount of generated entanglement and allow the system to approach the bound more closely, but it cannot relax the universal threshold itself. In Sec.~\ref{sec:universality}, we derive an initial state independent separability-preservation condition and show
that the same bound gives the universal threshold for two-mode Gaussian
separable inputs.

\begin{figure*}[tb]
    \includegraphics[width=0.45\textwidth]{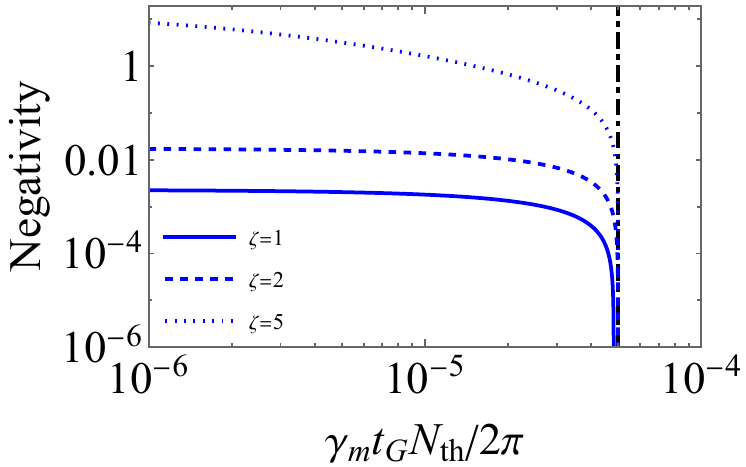}
    \includegraphics[width=0.45\textwidth]{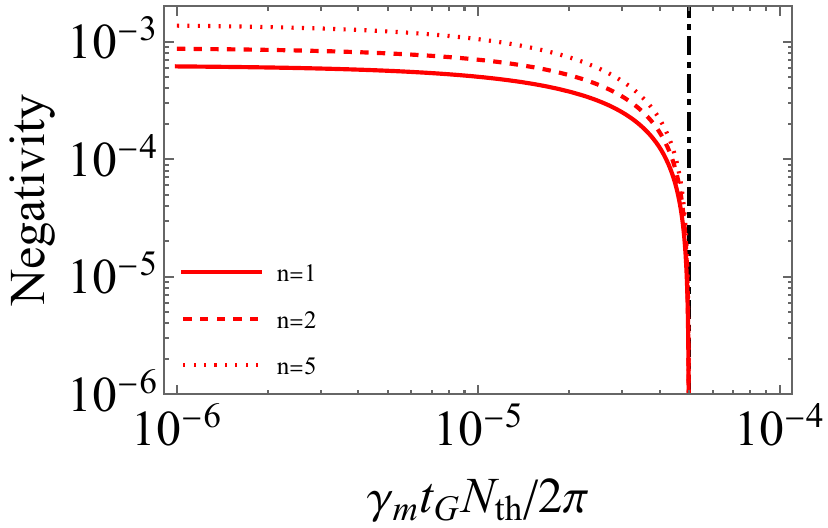}
    \caption{Behavior of entanglement negativity with the thermal noise. We fix the gravitational coupling $g_Gt_G/2\pi=10^{-4}$. In the left panel, the solid, dashed, and dotted curves correspond to the initial double-squeezed state with a squeezing parameter $\zeta=1$, $\zeta=2$, and $\zeta=5$. In the right panel, the solid, dashed, and dotted curves correspond to the initial Fock-vacuum state with $n=1$, $n=2$, and $n=5$. The vertical black dashed-dotted line represents the value $g_Gt_G/4\pi$. Entanglement is generated in the region where condition~\eqref{gie} is satisfied.}
    \label{fig:nphth}
\end{figure*}

\subsection{Initial Fock state}
Next, we consider using Fock states for the ingoing optical field.
In this non-Gaussian case, covariance matrices cannot fully characterize entanglement, and we resort to the {\it positive partial-transpose (PPT)} criterion, which identifies entanglement by finding negative eigenvalues of the partially transposed density matrix of bipartite systems.

To specify the input quantum state, we note in Fig~\ref{fig:pulse} that there are a total of six input operators; from top to bottom, the first three are input optical field for mass A's second pulse, input optical field for mass A's first pulse, mass A itself, while the bottom three are for mass B and its optical fields.  In this paper, we will use 
$\ket{\Psi_{\rm in}}=\ket{0,0}_m\ket{n,0}_{p(1)}\ket{0,0}_{p(2)}$ where the mechanical modes and the second pulse fields are in the vacuum state, with first pulse of mass A in the $|n\rangle$ and first pulse of mass B in vacuum.
In the Heisenberg picture, the quantum state remains constant throughout the three evolution stages, therefore 
\begin{equation}
    \ket{\Psi_{\rm out}}=\frac{1}{\sqrt{n!}}(\hat \xi_{A,{\rm in}}^{(1)\dagger})^n\ket{0,0}_m\ket{0,0}_{p(1)}\ket{0,0}_{p(2)}
\end{equation}
This can be expressed in terms of creation operators of the outgoing fields if we use
\begin{align}
    \hat \xi_{A,{\rm in}}^{(1)}&=
    -\cos[g_Gt_G]\hat \xi_{A,{\rm out}}^{(2)}+i\sin[g_Gt_G]\hat \xi_{B,{\rm out}}^{(2)}\,,
\end{align}
which can be obtained by inverting Eqs.~\eqref{xiathermal} and \eqref{xibthermal}. This leads to the following density matrix:
\begin{align}
    \label{rhon}
    \rho_{\rm out}^{(2)}&=
    \sum_{k,k^\prime=0}^n\sqrt{\frac{(n!)^2}{k!(n-k)!k^\prime !(n-k^\prime)!}}
    (-1)^{n+k^\prime}\nonumber\\
    &\quad\times\left(\cos[g_Gt_G]\right)^{k+k^\prime}
    \left(i\sin[g_Gt_G]\right)^{2n-k-k^\prime}\nonumber\\
    &\quad\times\ket{k,n-k}\bra{k^\prime,n-k^\prime}.
\end{align}
The entanglement negativity is defined as the sum of the absolute values of the negative eigenvalues of the partial transposed reduced density matrix of the outgoing field, and is given by:
\begin{align}
    \begin{split}
        \mathcal N_n&=
        \frac{1}{2}
        \left(\sum_{k=0}^n\Lambda_k^{(n)}\right)^2-\frac{1}{2}\\
        \Lambda_k^{(n)}&=\sqrt{\binom{n}{k}}(|\cos[g_Gt_G]|)^k(|\sin[g_Gt_G]|)^{n-k}.
    \end{split}
\end{align}
This expression also shows that entanglement negativity is a monotonically nondecreasing function of the Fock number $n$ of the initial state. To see this, we define $\bm v_k=(|\cos[g_Gt_G]|\Lambda_{k-1}^{(n)},|\sin[g_Gt_G]|\Lambda_k^{(n)})^{\rm T}$. Using the triangle inequality, we obtain $\sum_k\Lambda_k^{(n+1)}=\sum_k||\bm v_k||\ge ||\sum_k\bm v_k||=\sum_k\Lambda_k^{(n)}$. Thus, $\mathcal N_{n+1}\ge\mathcal N_n$. The equality holds only when $g_Gt_G=m\pi/2$ with $m\in\mathbb Z$. In this case, we have $\mathcal N_n=0$. Therefore, in the nontrivial regime where gravity generates entanglement, increasing the input Fock number enhances the entanglement negativity.

Next, we consider the effect of thermal noise. We shall rewrite the system's quantum state in terms of its density matrix:
$\rho_{\rm in}=\ket{0,0}_m\bra{0,0}\otimes\ket{n,0}_{p(1)}\bra{n,0}\otimes\ket{0,0}_{p(2)}\bra{0,0}\otimes\rho_{\rm th}$, where $\rho_{\rm th}$ is the density matrix of the environment. The reduced density matrix of the {second-pulse output} optical modes can be written as
\begin{align}
    \rho_{{\rm out},{\rm th}}^{(2)}&=
    \int \frac{d^2\alpha_A}{\pi}\int \frac{d^2\alpha_B}{\pi}
    \chi_{\rm out}(\alpha_A,\alpha_B)\nonumber\\
    &\qquad\times\hat D_A^{(2)}(-\alpha_A)\hat D_B^{(2)}(-\alpha_B),
\end{align}
where $\hat D_j^{(2)}(\alpha_j)=e^{\alpha_j^*\hat \xi_{j,{\rm out}}^{(2)}-\alpha_j\hat \xi_{j,{\rm out}}^{(2)\dagger}}$ is the displacement operator of the second optical modes and $\chi_{\rm out}={\rm tr}\left[\rho_{\rm in}\hat D_A^{(2)}(\alpha_A)\hat D_B^{(2)}(\alpha_B)\right]$ is the characteristic function. We obtain the analytic expressions of the characteristic function and the reduced density matrix in the Fock basis in Appendix~\ref{apdx:thermal}.

\subsection{Amount of Entanglement}
We now present predictions for the amount of gravity-induced entanglement generated by the pulsed protocol, with both Gaussian and Fock input states and different levels of thermal noise, but in the absence of measurement loss. The purpose of this subsection is twofold. First, we show explicitly that nonclassical input states can amplify the generated entanglement. Second, we emphasize that this amplification does not relax the threshold condition derived above. The effects of imperfect readout and the question of how much entanglement is experimentally observable will be discussed separately in Sec.~\ref{sec:loss}.

Fig.~\ref{fig:nphth} shows the entanglement negativity as a function of the accumulated thermal decoherence scale $\gamma_m t_G N_{\rm th}/2\pi$ for a fixed gravitational phase $g_G t_G/2\pi=10^{-4}$. In the left panel, the input light fields during the first pulses are prepared in two squeezed vacuum states, while in the right panel the input light field for mass A during the first pulse is prepared as an $n$-Fock state.
The solid, dashed, and dotted curves correspond to $\zeta=1,2,5$ in the squeezed state case and to $n=1,2,5$ in the Fock state case, respectively. In both cases, increasing the nonclassicality of the input state increases the amount of entanglement once the system is already in the entangling regime. However, all curves disappear at the same thermal noise threshold, indicated by the vertical black dash-dotted line. This confirms that input squeezing or Fock state preparation amplifies the amount of generated entanglement, but does not move the fundamental boundary $g_G=2\gamma_m N_{\rm th}$.

\begin{figure}
    \centering
    \includegraphics[width=0.9\linewidth]{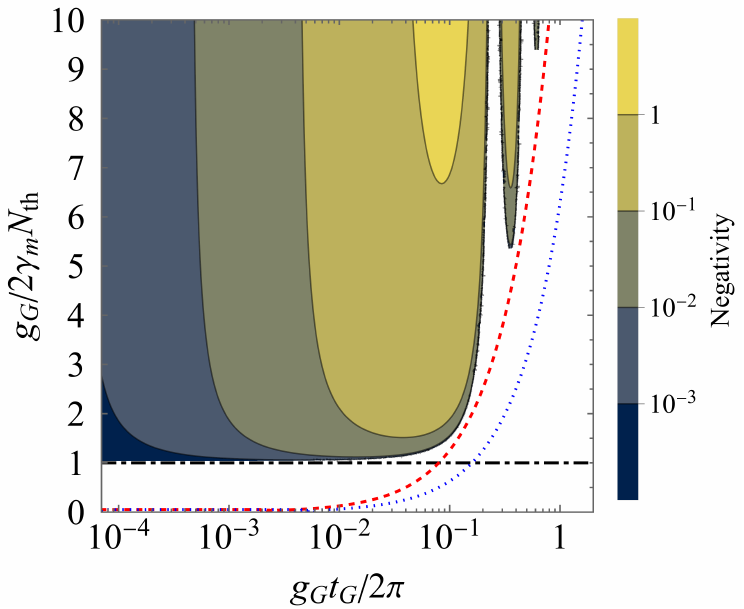}
    \caption{Entanglement negativity for the squeezed input with $\zeta=1$ in the dimensionless parameter space $g_Gt_G/2\pi$ and $g_G/2\gamma_mN_{\rm th}$. The horizontal black dashed-dotted line shows the universal threshold $g_G=2\gamma_mN_{\rm th}$. The red dash-dotted curve shows the entanglement-annihilating boundary $4\gamma_mt_GN_{\rm th}=1$, while the blue dotted curve to its right indicates the entanglement-breaking boundary $2\gamma_mt_GN_{\rm th}=1$.}
    \label{fig:timescale}
\end{figure}

To explore the broader parameter dependence, Fig.~\ref{fig:timescale} plots the entanglement negativity for constant input squeeze factor $\zeta=1$ in the dimensionless plane spanned by $g_G t_G/2\pi$ and $g_G/(2\gamma_m N_{\rm th})$. The horizontal black dash-dotted line denotes the universal threshold $g_G=2\gamma_m N_{\rm th}$. Below this line, thermal decoherence dominates over the coherent gravitational coupling, and no entanglement is generated.
The red dash-dotted and blue dotted curves denote the entanglement-annihilating boundary $4\gamma_m t_G N_{\rm th}=1$ and the entanglement-breaking boundary $2\gamma_m t_G N_{\rm th}=1$, respectively; both will be discussed in Sec.~\ref{sec:EBEA}. The physically relevant lossless entangling region therefore lies above the universal threshold and before the accumulated thermal noise reaches the entanglement-annihilating scale. Within this region, increasing $g_G t_G$ initially enhances the generated entanglement because the gravitational beam-splitter interaction has more time to accumulate. At larger interaction times, however, thermal noise also accumulates and eventually suppresses the entanglement.

\section{Entanglement bounds from thermal noise}
\label{sec:entanglement bound}
The previous section showed that, while nonclassical input states can increase the amount of output entanglement,
they do not shift the boundary between entangling and non-entangling dynamics in the $(g_G ,\gamma_m N_{\rm th})$ parameter space. In this section, we clarify this limitation from the viewpoint of Gaussian channels. We first derive a separability-preservation bound, which determines when initially separable two-mode Gaussian states remain separable under the gravitational interaction and thermal noise. We then discuss the entanglement-annihilating and entanglement-breaking conditions, which characterize a stronger regime where thermal noise destroys entanglement even for initially entangled states. These results show that thermal decoherence imposes two distinct constraints: the coherent gravitational coupling must exceed the thermal decoherence rate in order to generate entanglement from separable inputs, while the accumulated thermal noise during the interaction time must remain below an order-one phonon scale in order for any entanglement to survive.

\subsection{Universal Separability-Preservation Bound}
\label{sec:universality}
We first derive a universal condition for generating gravity-induced Gaussian entanglement in presence of Markovian thermal noise, for two oscillators with the same resonant frequency, with their gravitational interaction treated with Rotating-Wave Approximation (RWA) as done in this paper. A related form of this entanglement bound was originally discussed in Refs.~\cite{kafri2013,kafri2014} in the context of whether gravitational interactions can be implemented by classical channels and classical interactions.
Here, we revisit this  condition from the perspective of quantum gravity and environmental thermal noise. A more general discussion will be presented in Ref.~\cite{alfred2026}.

\noindent \textbf{Theorem: separability-preservation bound}.
In the regime $\gamma_mt_G\ll g_Gt_G\ll1$,  all initially separable two-mode Gaussian states remain separable under the RWA evolution if and only if 
\begin{align}
    g_G \le 2\gamma_m N_{\rm th}.
    \label{separability_bound_simple}
\end{align}
Equivalently, gravity-induced entanglement can be generated by preparing some initial state if and only if  $g_G > 2\gamma_m N_{\rm th}$.

\textbf{Proof}.
For Gaussian states, separability implies the inequality \cite{duan2000,simon2000}
\begin{align}
    \bm P\bm V\bm P+\bm \Sigma\ge0,
\end{align}
where $\bm P$ is the partial transpose matrix and $\bm \Sigma$ denotes the symplectic form. For two-mode Gaussian states, this condition is necessary and sufficient. In this case, $\bm P={\rm diag}\{1,1,1,-1\}$ and $\bm \Sigma=(-\bm \sigma_y)\oplus(-\bm \sigma_y)$ with the Pauli matrix $\bm \sigma_y$. Using the output covariance matrix $\bm V_{\rm out}$ in Eq.~\eqref{double squeezed CM}, the above inequality is equivalent to
\begin{align}
\label{transformed PPT}
    \bm V_{\rm in}+\bm M^{-1}\bm P\bm \Sigma\bm P(\bm M^{-1})^{\rm T}+\bm M^{-1}\bm V_{\rm th}(\bm M^{\rm T})^{-1}\ge0.
\end{align}
We denote the left-hand side by
\begin{align}
    \tilde{\bm V}=\bm V_{\rm in}+\bm \Sigma_{\rm PT}+\bm \Omega
\end{align}
where $\bm \Sigma_{\rm PT}=\bm P\bm \Sigma\bm P$ and
\begin{align}
    \bm \Omega&=
    \left(\begin{array}{cccc}
    \Omega_1 & i\Omega_2 & i\Omega_3 & 0 \\
    -i\Omega_2 & \Omega_1 & 0 & i\Omega_3 \\
    -i\Omega_3 & 0 & \Omega_1 & -i\Omega_2 \\
    0 & -i\Omega_3 & i\Omega_2 & \Omega_1
    \end{array}\right),
    \label{Omegamat}
\end{align}
with $\Omega_1=4e^{\gamma_mt_G}N_{\rm th}\sinh[\gamma_mt_G]$, $\Omega_2=e^{2\gamma_mt_G}\cos[2g_Gt_G]-1$, and $\Omega_3=e^{2\gamma_mt_G}\sin[2g_Gt_G]$.
Since the initial state is a separable two-mode Gaussian state, it satisfies $\bm{V}_{\rm in}+\bm{\Sigma}_{\text{PT}} \geq 0$. Therefore, if $\bm \Omega\ge0$, then $\tilde{\bm V}\ge0$ for any initial separable two-mode Gaussian state. Hence, $\bm \Omega\ge0$ is a sufficient condition for separability preservation. The matrix $\bm \Omega$ is positive semidefinite if and only if $\Omega_1 \ge \sqrt{\Omega_2^2+\Omega_3^2}$. In the regime $\gamma_mt_G\ll g_Gt_G\ll1$, this inequality implies $g_G\le2\gamma_mN_{\rm th}$. This proves that Eq.~\eqref{separability_bound_simple} is sufficient for separability preservation.

Moreover, for $\Omega_2\ge0$, the condition $\bm \Omega\ge0$ is also necessary. We show the details of the proof in Appendix \ref{apdx: necessary}. In the regime $\gamma_mt_G\ll g_Gt_G\ll1$, we have $\Omega_2\ge0$. Hence, the condition $2\gamma_mN_{\rm th}\ge g_G$ is necessary and sufficient condition for separability preservation $\square$.

The form of the criterion can be physically interpreted as requiring that the rate of gravitational coupling must be greater than the thermal decoherence rate. We note that the condition derived here is a criterion for the preservation of separability. Equivalently, if $\bm \Omega\not\ge0$, it is no longer guaranteed that an initially separable state will remain separable during evolution. This does not mean that the state must become entangled for every initial state; rather, it means that entanglement becomes possible for some states. Indeed, depending on the choice of the initial state, there exist states that remain separable even when this inequality is violated.

\subsection{Entanglement Breaking and Annihilating}
\label{sec:EBEA}
So far, we considered initially separable states and derived the universal condition under which the state remains separable. Another point of view is to consider arbitrary Gaussian input states, including the entangled ones, and and ask thresholds of thermal decoherence that destroy entanglement. The relevant criteria in this case are discussed in terms of the entanglement-annihilating (EA) and entanglement-breaking (EB) conditions \cite{horodecki2003,holevo2008,moravcikova2010,filippov2014}.  An EA channel is a channel acting on the system of interest such that, for any initial state of that system, the output state is separable. An EB channel is a channel acting on the system of interest such that, even when an ancillary system is included, any initial state of the total system evolves into a separable state with respect to the bipartition between the system of interest and the ancilla. We show that, unlike the separability-preservation condition discussed above, these conditions are independent of the gravitational interaction, but are determined by the thermal decoherence rate, as well as interaction time. They provide complementary constraints on experimental protocols.

We first discuss the EA condition. For continuous variable systems, Ref.~\cite{filippov2014} showed that the EA condition can be obtained by taking as the input state the infinite-squeezing limit of a two-mode squeezed state, which is the maximally entangled state. We therefore consider the following covariance matrix for the initial state:
\begin{align}
    \bm V_{\rm in}^{\rm max}&=
    \left(\begin{array}{cc}
    \cosh2r \bm 1 & \sinh 2r \bm \sigma_z \\
    \sinh 2r \bm \sigma_z & \cosh 2r \bm 1
    \end{array}\right),
\end{align}
where $\bm \sigma_z$ is the Pauli matrix and $r$ is the squeezing parameter. Using the evolution in Eq.~\eqref{double squeezed CM}, we obtain the following condition from the requirement that the minimum symplectic eigenvalue after partial transposition satisfy $\nu_-\ge1$:
\begin{align}
    4\gamma_mt_GN_{\rm th}\ge1,
\end{align}
where we used $r\to \infty$.
Therefore, if this condition is satisfied, the output state becomes separable for any initial Gaussian state. The dimensionless quantity $\gamma_mt_GN_{\rm th}$ characterizes the amount of thermal noise injected into the mechanical mode during the interaction time. This inequality implies that once thermal noise injects $\mathcal O(1)$ phonon over the interaction time, the channel becomes entanglement annihilating.

Next, we consider the EB condition. A channel $\Phi_{AB}$ is entanglement breaking if, for an arbitrary ancillary system C, the evolved state $(\Phi_{AB}\otimes 1)(\rho_{AB-C})$ is separable for any initial state $\rho_{AB-C}$. Since the dynamics in our model is Gaussian, we can apply the EB criterion for Gaussian channels \cite{holevo2008}. For the covariance-matrix evolution in Eq.~\eqref{double squeezed CM}, Ref.~\cite{holevo2008} showed that a two-mode Gaussian channel is EB if and only if there exist positive semidefinite matrices $\bm V_1,\bm V_2$ such that
\begin{align}
    \bm V_{\rm th}=\bm V_1+\bm V_2,\quad
    \bm V_1+\bm \Sigma\ge0,\quad
    \bm V_2+\bm M\bm \Sigma \bm M^{\rm T}\ge0.
\end{align}
where $\bm \Sigma=(-\bm \sigma_y)\oplus(-\bm \sigma_y)$ with the Pauli matrix $\bm \sigma_y$.
In our case, the noise is isotropic $\bm V_{\rm th}=2(1-e^{-2\gamma_mt_G})N_{\rm th}$, and therefore we can choose an isotropic decomposition $\bm V_1=\alpha \bm 1$ and $\bm V_2=(2(1-e^{-2\gamma_mt_G})N_{\rm th}-\alpha)\bm 1$. Then, the conditions reduce to the scalar constraints
\begin{align}
    \alpha\ge1,\quad
    2(1-e^{-2\gamma_mt_G})N_{\rm th}-\alpha\ge e^{-2\gamma_mt_G}.
\end{align}
so that a feasible $\alpha$ exists if and only if $2\tanh[\gamma_mt_G]N_{\rm th}\ge1$. Assuming the weak dissipation $\gamma_mt_G\ll1$, this becomes
\begin{align}
    2N_{\rm th}\gamma_mt_G\ge1.
\end{align}
Hence, once $2\gamma_mt_GN_{\rm th}\ge1$, the output state is separable from any ancilla, even if the initial joint state is entangled. In the present model, however, no ancillary system is included. Therefore, the relevant boundary is given by the EA condition, while the EB bound lies further to the right, as shown in Fig.~\ref{fig:timescale}.

\section{Measurement loss effects}
\label{sec:loss}
We now turn from the generation of mechanical entanglement to its observable optical readout.  In the idealized discussion of Sec.~\ref{sec:twoopto}, the entanglement generated during the gravitational interaction is assumed to be perfectly swapped back to the second optical pulses and perfectly measured.  In practice, however, the output temporal modes suffer from losses and mode mismatch before they are reconstructed.  We collect these imperfections into an effective readout efficiency $\eta$. Thus, $\eta$ should be interpreted as the total efficiency with which the second-pulse output mode entering the covariance matrix is recovered in the experimentally reconstructed optical mode.  It therefore includes not only the quantum efficiency of the photodetectors, but also optical propagation loss, imperfect mode matching, finite homodyne visibility, and any other loss that can be modeled as a local Gaussian attenuation channel.  Additional excess noise in the readout stage can be incorporated by replacing the vacuum environment below by a thermal environment; in this section we focus on the pure-loss case in order to isolate the effect of inefficient readout.

\begin{figure}[tb]
    \centering
    \includegraphics[width=0.9\linewidth]{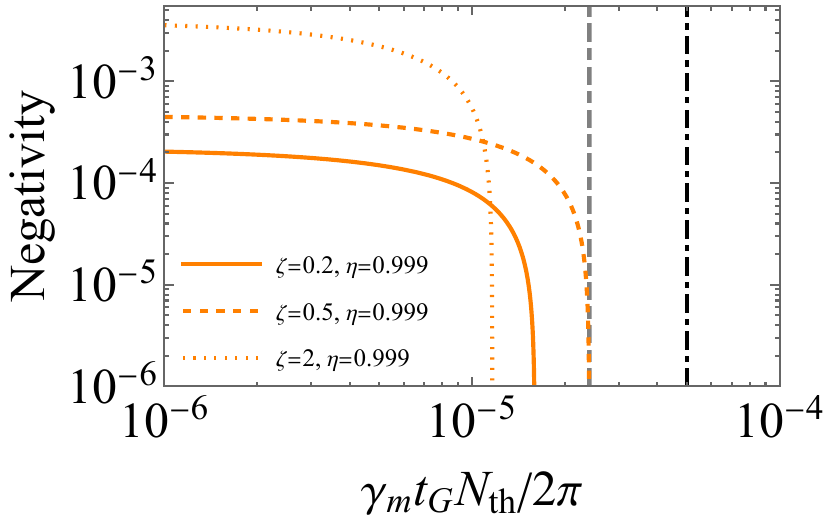}
    \caption{Behavior of entanglement negativity with measurement loss. We fix the gravitational coupling $g_Gt_G/2\pi=10^{-4}$.
    The solid, dashed, and dotted curves correspond to a fixed transmittance $\eta = 0.999$ with squeezing values $\zeta = 0.2$, $\zeta = 0.5$, and $\zeta = 5$, respectively.
    The vertical black dash-dotted line indicates the lossless universal bound $g_G=2\gamma_m N_{\rm th}$, while the gray dash-dotted line to its left indicates the lossy bound given by Eq.~\eqref{lossybound}. Unlike the lossless case, the lossy bound is not saturated in the large-squeezing limit. Instead, for a given measurement efficiency, there exists a finite input squeezing that approaches the loss-modified bound most closely, as illustrated here by the curve with $\zeta=0.5$.}
    \label{fig:loss negativity}
\end{figure}

\subsection{Mathematical Model}
When the final output modes experience measurement loss, this process is modeled via the input-output relation on annihilation operators~\cite{Brady_2024} as
\begin{equation}
    \hat{\xi}^{(2),\text{post-loss}}_{j,\text{out}} = \sqrt{\eta_j}\hat{\xi}^{(2)}_{j,\text{out}}+\sqrt{1-\eta_j}\hat{e}_j
\end{equation}
where $0\leq \eta_j\leq 1$ is the transmittance of the output modes and $\hat{e}_j$ corresponds to the environment mode in a Gaussian thermal state with mean $\langle \hat{e}_j^{\dagger}\hat{e}_j\rangle = \bar{n}_j$ photons. Since loss is a local Gaussian channel and cannot increase entanglement, it suffices to consider a vacuum environment with $\bar{n}_j = 0$ (this is also known as pure-loss). We will only study its effect on Gaussian input states as it already suffices to demonstrate the key effects of loss. Converting from $\hat{\xi}$ to $\hat{X},\hat{Y}$ using the same definitions as before, we derive the following input-output relation on covariance matrices (also see Appendix \ref{apdx: post loss CM}):
\begin{align}
    \bm V_{\rm post-loss}&=
    \eta \bm V_{\rm pre-loss}+(1-\eta)\bm 1
\end{align}
where $\bm V_{\rm pre-loss}$ and $\bm V_{\rm post-loss}$ are the covariance matrix without and with measurement loss. We fix $\eta_{A} = \eta_{B} =\eta$ for simplicity.

The loss channel modifies the separability-preservation condition. Using the same method as in Sec.~\ref{sec:universality}, but applying the PPT condition to $\bm V_{\rm post-loss}$, we obtain the separability-preservation condition, in the regime $\gamma_m t_G\ll g_Gt_G\ll1$,
\begin{align}
    4\eta\gamma_m t_GN_{\rm th}
    \ge
    \sqrt{(1-\eta)^2+4\eta(g_Gt_G)^2}-(1-\eta).
    \label{lossybound}
\end{align}
The detailed derivation is shown in Appendix~\ref{apdx: necessary}.
This condition reduces to the same condition $g_G\le 2\gamma_mN_{\rm th}$ in the lossless limit $\eta=1$. For $\eta<1$, however, the condition becomes explicitly dependent on the interaction time $t_G$. Physically, measurement loss adds a finite amount of vacuum noise to the reconstructed output mode, and therefore a state that would be entangled before readout can become separable after inefficient measurement, unless the level of entanglement exceeds a loss-dependent threshold.

\begin{figure*}
    \centering
    \includegraphics[width=0.7\linewidth]{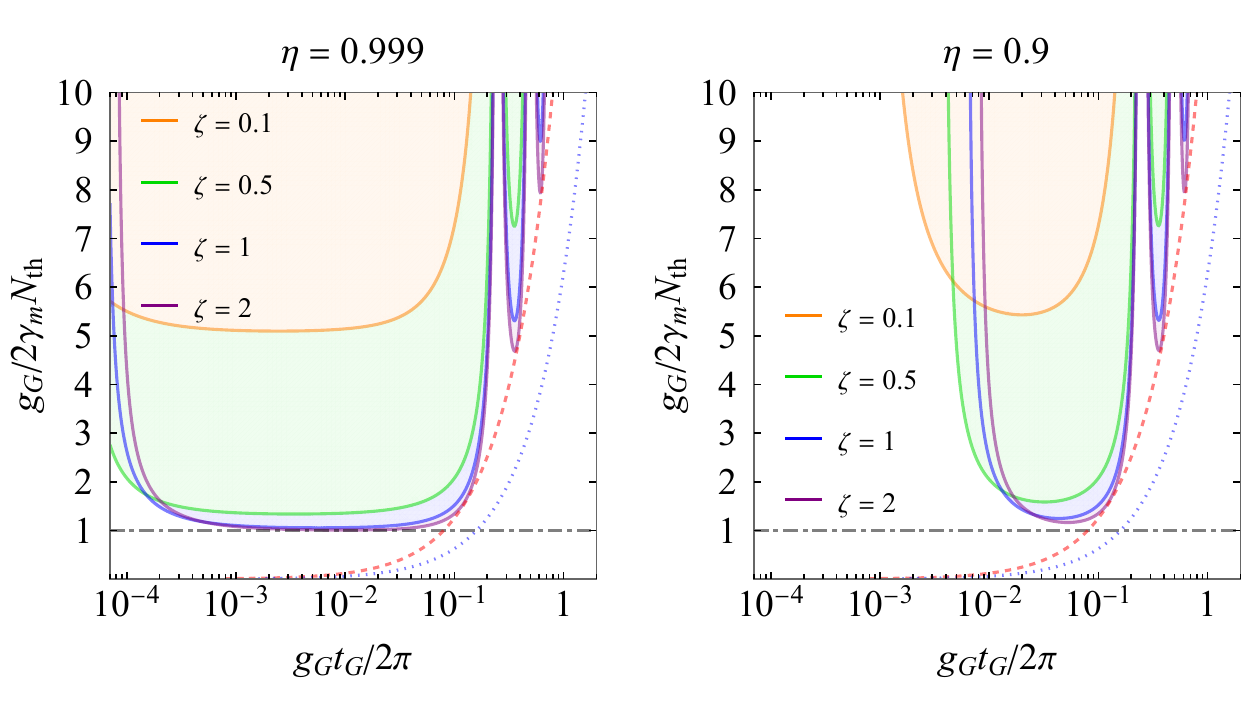}
    \includegraphics[width=0.7\linewidth]{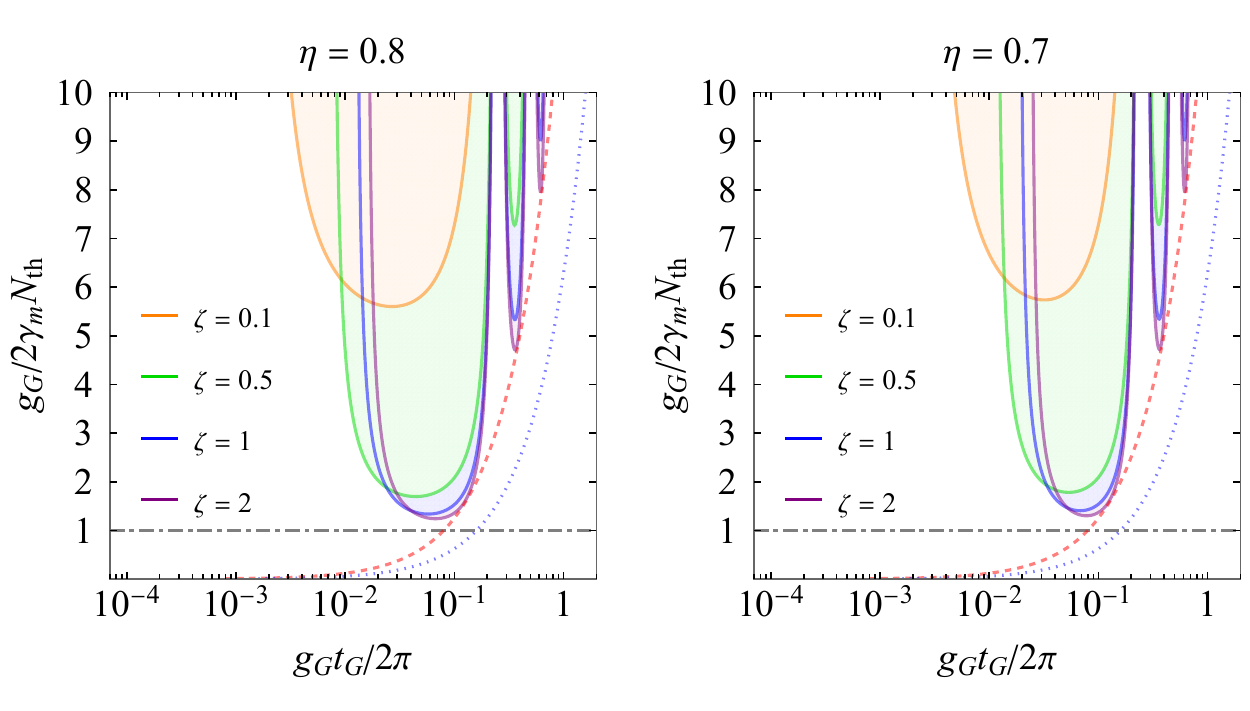}
    \caption{Entanglement region in the $(g_Gt_G/2\pi,\,g_G/2\gamma_mN_{\rm th})$ plane for several squeezing parameters and measurement efficiencies. The four panels correspond to $\eta=0.999,0.9,0.8,0.7$, from top left to bottom right. The black dash dotted line, red dashed curve, and blue dotted curve are the same reference lines as in Fig.~\ref{fig:timescale}. The solid curves denote the boundaries where the negativity vanishes for $\zeta=0.1,0.5,1,2$, shown in orange, green, blue, and purple, respectively. The shaded region above each boundary indicates the parameter region where entanglement is generated.}
    \label{fig:optimal}
\end{figure*}

\begin{figure*}
    \centering
    \includegraphics[width=0.45\linewidth]{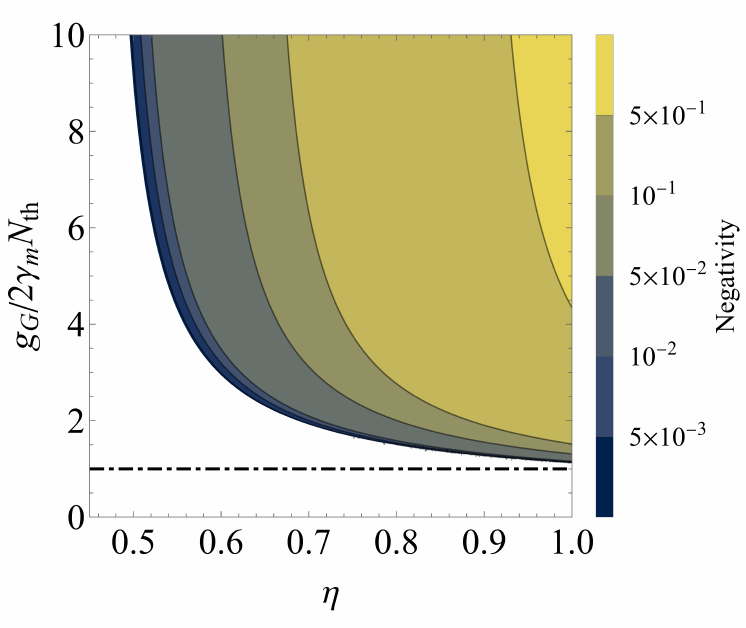}
    \includegraphics[width=0.45\linewidth]{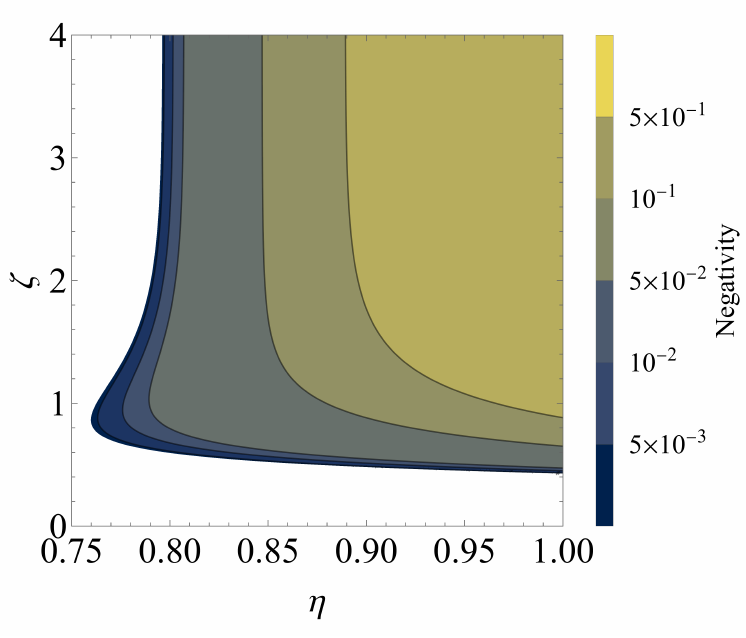}
    \caption{
    Effect of measurement loss on the entanglement negativity. In both panels, we fix $g_Gt_G/2\pi=3.4\times10^{-2}$. Left: the negativity as a function of the measurement loss $\eta$ and the ratio between the gravitational coupling and thermal decoherence, $g_G/2\gamma_mN_{\rm th}$, for a fixed input squeezing $\zeta=1$. The black dash-dotted line indicates the universal bound $g_G/2\gamma_mN_{\rm th}=1$ in the absence of measurement loss. As $\eta$ decreases, the requirement on thermal decoherence becomes more stringent; for a fixed gravitational coupling, $\gamma_mN_{\rm th}$ must be reduced further. Right: the negativity as a function of $\eta$ and the input squeezing parameter $\zeta$, for a fixed ratio $g_G/2\gamma_mN_{\rm th}=1.6$. In contrast to the lossless case, the negativity has an optimal region around $\zeta\simeq0.8$, reflecting the existence of an optimal squeezing that approaches the lossy bound shown in Fig.~\ref{fig:loss negativity}. For larger squeezing, $\zeta\gtrsim2$, the negativity becomes almost insensitive to further increasing $\zeta$.}
    \label{fig:sqlossy}
\end{figure*}

\subsection{Experimental Feasibility of Detectable Gravity-Induced Entanglement}
Let us now explore the consequences of measurement losses on detectable levels of entanglement.  Figure \ref{fig:loss negativity} shows this effect for a squeezed input state. We fix the gravitational phase to $g_Gt_G/2\pi=10^{-4}$ and the measurement efficiency to $\eta=0.999$, and plot the entanglement negativity as a function of the accumulated thermal decoherence scale $\gamma_m t_GN_{\rm th}/2\pi$. The solid, dashed, and dotted curves correspond to $\zeta=0.2$, $\zeta=0.5$, and $\zeta=2$, respectively. The vertical black dash-dotted line shows the lossless universal bound $g_G=2\gamma_mN_{\rm th}$, while the gray dash-dotted line to its left shows the lossy bound in Eq.~\eqref{lossybound}. The shift of the bound to the left means that, once measurement loss is present, the allowed thermal decoherence must be smaller than in the ideal lossless case. An important difference from the lossless case is the role of input squeezing. Without measurement loss, increasing the squeezing allows the squeezed-input protocol to approach the lossless universal bound. With measurement loss, however, large squeezing also increases the sensitivity to the vacuum noise injected by the loss channel. As a result, the large-squeezing limit no longer saturates the relevant bound. Instead, for a given value of $\eta$, there is a finite input squeezing that approaches the lossy bound most closely. In Fig.~\ref{fig:loss negativity}, this behavior is illustrated by the curve with $\zeta=0.5$, which reaches closer to the gray lossy-bound line than the more strongly squeezed case $\zeta=2$.

Figure~\ref{fig:optimal} shows how measurement loss and input squeezing modify the entanglement region. Each solid curve denotes the boundary where the negativity vanishes, and entanglement is generated in the shaded region above it. The black dash dotted line, red dashed curve, and blue dotted curve are the same reference lines as in Fig.~\ref{fig:timescale}. In each panel, the squeezing parameter is varied as $\zeta=0.1,0.5,1,2$, and the four panels correspond to $\eta=0.999,0.9,0.8,0.7$.
In the absence of measurement loss, the lower bound on the interaction time is essentially set by the validity of the RWA, $\omega_m t_G \gg 1$. In the presence of measurement loss, however, there appears an optimal interaction time at which the requirement on thermal noise is minimized. The universal bound itself also becomes $\eta$ dependent, as shown in Eq.~\eqref{lossybound}. For smaller $\eta$, the short-time entangling region is suppressed, so that a longer interaction time and a lower level of thermal noise are required to observe entanglement. Moreover, increasing the input squeezing reduces the minimum thermal noise requirement and brings the boundary closer to the universal bound, although this comes at the cost of shifting the entangling region to longer interaction times. Thus, there is a tradeoff between the amount of input squeezing and the interaction time required to observe entanglement. For the parameter range shown here, the thermal noise requirement is minimized at an intermediate interaction time, typically between $10^{-2}$ and $10^{-1}$ in $g_Gt_G/2\pi$.

Motivated by this intermediate optimal timescale, we next examine the amount of entanglement in Fig.~\ref{fig:sqlossy}, fixing the accumulated gravitational phase to $g_Gt_G/2\pi=3.4\times10^{-2}$.
In the left panel, the input squeezing is fixed to $\zeta=1$, while the ratio $g_G/2\gamma_mN_{\rm th}$ is varied. The black dash-dotted line denotes the lossless universal bound, 
$g_G/2\gamma_mN_{\rm th}=1$. As the measurement loss $\eta$ decreases, the same amount of negativity requires a larger value of $g_G/2\gamma_mN_{\rm th}$. Since $g_G$ is essentially fixed by the gravitational constant and the mass density scale $M/d^3$, this should be interpreted primarily as a more stringent requirement on suppressing thermal decoherence, 
i.e., reducing $\gamma_mN_{\rm th}$. The right panel shows the dependence on the input squeezing $\zeta$ for a fixed ratio $g_G/2\gamma_mN_{\rm th}=1.6$. Unlike the lossless case, where increasing the squeezing monotonically approaches the ideal bound, the lossy case exhibits an optimal squeezing around $\zeta\simeq0.8$. 
When measurement loss is present, excessive squeezing also amplifies the noise contribution associated with the inefficient readout, and therefore the best observable negativity is obtained at a finite squeezing. For $\zeta\gtrsim2$, the negativity changes only weakly with further increasing $\zeta$, indicating that the measurement loss already limits the accessible entanglement in this regime.

We now connect these results to experimental requirements. In the absence of measurement loss, the universal threshold, the requirement of avoiding the entanglement-annihilating regime, and the RWA condition can be summarized as
\begin{align}
    Q_m>\frac{k_BT}{\hbar g_G}, \qquad \omega_m^{-1}\ll t_G<\frac{\hbar Q_m}{2k_BT},
    \label{eq:exp requirement}
\end{align}
where $Q_m=\omega_m/2\gamma_m$. The measurement loss discussed above adds an additional restriction: even when the generated mechanical entanglement satisfies the lossless condition, the reconstructed output optical state can become separable unless the accumulated gravitational phase, the input squeezing, and the measurement efficiency are sufficiently large.

To illustrate the scale of the required parameters, we use
\begin{align}
    &\frac{g_G}{2\gamma_mN_{\rm th}} \simeq 1.6 \left(\frac{M/d^3}{20~{\rm g}/{\rm cm}^3}\right) \left(\frac{10^{-15}~{\rm Hz}}{2\gamma_m/2\pi}\right) \left(\frac{1~{\rm mK}}{T}\right),\\
    &\frac{g_Gt_G}{2\pi} \simeq 3.4\times10^{-2} \left(\frac{M/d^3}{20~{\rm g}/{\rm cm}^3}\right) \left(\frac{t_G}{10^4~{\rm sec}}\right) \left(\frac{10~{\rm mHz}}{\omega_m/2\pi}\right).
\end{align}
For these representative values, the system lies above the lossless universal threshold, $g_G/2\gamma_mN_{\rm th}=1$. Moreover, as shown in Fig.~\ref{fig:sqlossy}, the post-loss entanglement survives for moderate input squeezing and sufficiently high measurement efficiency. In particular, around the optimal squeezing region $\zeta\simeq0.8$, entanglement remains observable for $\eta\gtrsim0.75$, and the entanglement negativity can exceed $\mathcal N_G\sim10^{-3}$.

This value is above the rough resolution scale inferred from existing continuous variable entanglement experiments, where logarithmic negativities of order $10^{-1}$ have been reconstructed with uncertainties at the level of a few percent~\cite{palomaki2013,junxin2020}. Although those experiments were performed in different frequency regimes and with different optomechanical or optical couplings from the present proposal, they suggest that, in continuous variable platforms, an experimentally relevant resolution for entanglement negativity is roughly at the $10^{-3}$--$10^{-2}$ scale. Therefore, the present analysis provides a set of target parameters for reaching this observable entanglement scale after measurement loss.

\section{Summary and conclusion}
\label{sec:summary}
We have studied a pulsed optomechanical protocol for amplifying and reading out gravity-induced entanglement. In this protocol, a first red-detuned optical pulse swaps a nonclassical optical input state onto each mechanical oscillator, the two oscillators then interact through their mutual gravitational coupling for a duration $t_G$, and a second optical pulse maps the final mechanical state back onto outgoing optical fields. This setup provides a direct way to use optical state engineering to prepare nonclassical mechanical states, while still allowing the final entanglement to be measured optically.

We first showed that nonclassical input states can substantially increase the amount of gravity-induced entanglement. For Gaussian inputs, appropriately oriented single-mode squeezed states are converted into two-mode entanglement by the beam-splitter-like gravitational interaction. For non-Gaussian inputs, a Fock-vacuum input also enhances the negativity, with larger Fock number leading to a larger generated entanglement in the nontrivial entangling regime. These results demonstrate that pulsed optomechanics can act as an entanglement amplifier: once the system is already in the entangling regime, increasing the nonclassicality of the input state can increase the output negativity.

At the same time, we found that this amplification does not relax the fundamental threshold for entanglement generation. In the high-temperature regime, the onset of gravity-induced entanglement is governed by the competition between the coherent gravitational coupling and mechanical thermal decoherence. For the symmetric two-mode system, this gives the condition \eqref{gie}. For squeezed Gaussian inputs, this condition is approached in the large squeezing limit in the absence of measurement loss, but it cannot be surpassed. We then proved the corresponding separability-preservation condition for arbitrary separable two-mode Gaussian inputs under the rotating-wave approximation. Thus, although the choice of input state can amplify the amount of entanglement, it cannot shift the boundary between entangling and non-entangling dynamics. The proof of the universal condition presented here assumes two-mode Gaussian states, and for non-Gaussian states we have only demonstrated the result in the case of Fock states. A more general treatment will be presented in Ref. \cite{alfred2026}.

We also derived entanglement-annihilating and entanglement-breaking conditions for the Gaussian channel generated by the gravitational interaction and thermal noise. These conditions describe a different limitation from the threshold condition above. While the threshold $g_G>2\gamma_mN_{\rm th}$ determines whether the gravitational interaction can overcome thermal decoherence and generate entanglement from an initially separable state, the entanglement-annihilating and entanglement-breaking conditions characterize regimes in which accumulated thermal noise destroys all output entanglement, even when one allows more general input states. We showed that these regimes appear when the thermal noise injects an $O(1)$ number of phonons during the interaction time, and that the corresponding conditions are independent of the gravitational coupling strength.

We further analyzed the effect of measurement loss on the outgoing optical fields. Measurement loss was modeled as a local pure-loss channel acting after the second pulse. Since this channel is local, it cannot create entanglement, but it can destroy the entanglement generated during the gravitational interaction before it is observed in the output light. We derived a loss-modified separability-preservation condition and showed that, unlike the lossless bound, the observable bound depends explicitly on the interaction time and on the measurement efficiency. In particular, the large-squeezing limit is no longer optimal when measurement loss is present. Instead, for a fixed measurement efficiency, there is a finite input squeezing that most closely approaches the lossy bound and maximizes the observable negativity.

\section{Acknowledgments}
D.M.\ is supported by the JSPS Overseas Research Fellowships and the Keck Foundation. 
Y.C.\ and A.L.\ are supported by the Simons Foundation (Award Number 568762).

\onecolumngrid

\appendix
\section{Derivation of the Hamiltonian and output operators in two optomechanical systems}
\label{apdx:Hamiltonian}
Here, we consider the two pulsed optomechanical systems interacting via gravity. The Hamiltonian is
\begin{align}
    \hat H=&
    \sum_{j=A,B}\left[\frac{\hat p_j^2}{2M_j}+\frac{1}{2}M_j\omega_{mj}^2\hat x_j^2
    +\hbar\omega_c\hat a_j^\dagger\hat a_j
    +i\hbar E_j(t)\left(e^{-i\omega_{Lj} t}\hat a_j^\dagger-e^{i\omega_{Lj}t}\hat a_j\right)\right]
    \nonumber\\
    &-\hbar \frac{\omega_{cA}}{\ell_A}\hat x_A\hat a_A^\dagger\hat a_A+\hbar\frac{\omega_{cB}}{\ell_B}\hat x_B\hat a_B^\dagger\hat a_B
    +\hat H_G,
\end{align}
where the summation part describes the Hamiltonian of the mechanical oscillation, the cavity modes, and the external driving lasers. The first two terms in the second line are the optomechanical interaction and the last term is the gravitational interaction given by
\begin{align}
    \hat H_G&=-\frac{GM_AM_B}{d-\hat x_A+\hat x_B}
    \sim
    -\frac{GM_AM_B}{d^3}(\hat x_A-\hat x_B)^2\nonumber\\
    &=-\frac{1}{2}M_A\omega_{AB}^2\hat x_A^2
    -\frac{1}{2}M_B\omega_{BA}^2\hat x_B^2
    +\frac{2GM_AM_B}{d^3}\hat x_A\hat x_B,
\end{align}
where $G$ is the gravitational constant, $d$ is the distance between two mirrors, and $\omega_{AB}^2=2GM_B/d^3$ and $\omega_{BA}^2=2GM_A/d^3$ are gravitational frequencies that merely shift the local mechanical frequencies.
In the following, we focus only on the last quadratic term proportional to $\hat x_A\hat x_B$: the other terms only renormalize the local mechanical frequencies and do not couple the two mirrors, while this cross term is the sole interaction responsible for entanglement generation between the two systems.
Introducing the phonon operators 
$\hat b_j=\sqrt{M_j\omega_{mj}/2\hbar}\hat x_J+i/\sqrt{\hbar M_j\omega_{mj}}\hat p_j$ and $\hat b_j^\dagger=\sqrt{M_j\omega_{mj}/2\hbar}\hat x_J-i/\sqrt{\hbar M_j\omega_{mj}}\hat p_j$,
the Hamiltonian in the rotating frame is
$\hat H_R=\hat H_0+\hat H_{\rm int}$, where
\begin{align}
    \hat H_0&=
    \sum_{j=A,B}\left[\hbar\omega_{mj}\hat b_j^\dagger\hat b_j
    +\hbar\Delta_j\hat a_j^\dagger\hat a_j\right]\\
    \hat H_{\rm int}=
    &-\hbar g_{A0}(\hat b_A+\hat b_A^\dagger)(\bar a_A^*\hat a_A+\bar a_A\hat a_A^\dagger)
    +\hbar g_{B0}(\hat b_B+\hat b_B^\dagger)(\bar a_B^*\hat a_B+\bar a_B\hat a_B^\dagger)
    +\hbar g_G(\hat b_A+\hat b_A^\dagger)(\hat b_B+\hat b_B^\dagger),
\end{align}
with optomechanical couplings $g_{j0}=\omega_{cj}x_{\rm zpj}/\ell_j$ with the zero point fluctuation $x_{\rm zpj}=\sqrt{\hbar/2M_j\omega_{mj}}$, gravitational coupling $g_G=G\sqrt{M_AM_B}/d^3\sqrt{\omega_{mA}\omega_{mB}}$, and detuning $\Delta_j=\omega_{Lj}-\omega_{cj}$.
In the interaction picture, the Hamiltonian is
\begin{align}
    \hat H_I =&
    -\hbar g_A(t)(e^{-i\omega_{mA}t}\hat b_A+e^{i\omega_{mA}t}\hat b_A^\dagger)(e^{i\Delta_At}\hat a_A+e^{-i\Delta_A t}\hat a_A^\dagger)
    +\hbar g_B(t)(e^{-i\omega_{mB}t}\hat b_B+e^{i\omega_{mB}t}\hat b_B^\dagger)(e^{i\Delta_Bt}\hat a_B+e^{-i\Delta_Bt}\hat a_B^\dagger)\nonumber\\
    &
    +\hbar g_G(e^{-i\omega_{mA}t}\hat b_A+e^{i\omega_{mA}t}\hat b_A^\dagger)(e^{-i\omega_{mB}t}\hat b_B+e^{i\omega_{mB}t}\hat b_B^\dagger),
\end{align}
where $g_j(t)=g_{j0}\bar a_j(t)$ is the time-dependent optomechanical coupling. Assuming the red-detuning $\Delta_j\sim-\omega_{mj}$ and $\omega_{mA}\sim\omega_{mB}$, we can drop the fast-phase term and derive
\begin{align}
    \hat H_I&=
    -\hbar g_A(t)(\hat b_A\hat a_A^\dagger+\hat b_A^\dagger\hat a_A)
    +\hbar g_B(t)(\hat b_B\hat a_B^\dagger+\hat b_B^\dagger\hat a_B)
    +\hbar g_G(\hat b_A\hat b_B^\dagger+\hat b_A^\dagger\hat b_B).
\end{align}

Let us consider the first pulse. The output operators after the first pulse is obtained by
\begin{align}
    \label{babb}
    &\hat b_A(t_P)=i\hat \xi_{A,{\rm in}}^{(1)},\quad
    \hat \xi_{A,{\rm out}}^{(1)}=i\hat b_{A,{\rm in}}\\
    &
    \hat b_B(t_P)=-i\hat \xi_{B,{\rm in}}^{(1)},\quad
    \hat \xi_{B,{\rm out}}^{(1)}=-i\hat b_{B,{\rm in}},
    \label{xiaxib}
\end{align}
where we assume the complete state swap.
$\hat \xi_{j,{\rm in}}^{(1)}$ and $\hat \xi_{j,{\rm out}}^{(1)}$ are the normalized temporal modes of the first pulse defined as
\begin{align}
    \hat \xi_{j,{\rm in}}^{(1)}&=
    \sqrt{\frac{2G_j}{e^{2G_jt_P}-1}}\int_0^{t_P}dse^{G_js}\hat a_{j,{\rm in}}(s)\\
    \hat \xi_{j,{\rm out}}^{(1)}&=
    \sqrt{\frac{2G_j}{1-e^{-2G_jt_P}}}\int_0^{t_P}dse^{-G_js}\hat a_{j,{\rm out}}(s),
\end{align}
where $G_j=g_j^2/\kappa$ and $[\hat \xi_{j,{\rm in}}^{(1)},\hat \xi_{j,{\rm in}}^{(1)\dagger}]=[\hat \xi_{j,{\rm out}}^{(1)},\hat \xi_{j,{\rm out}}^{(1)\dagger}]=1$.

Next, we consider the optomechanical interaction of the second pulse. The equations of motion are the same as the first pulse and we derive
\begin{align}
    \label{babbout}
    &\hat b_{A,{\rm out}}=
    i\hat \xi_{A,{\rm in}}^{(2)},\quad
    \hat \xi_{A,{\rm out}}^{(2)}=
    i\hat b_A(t_P+t_G)\\
    &\hat b_{B,{\rm out}}=
    -i\hat \xi_{B,{\rm in}}^{(2)},\quad
    \hat \xi_{B,{\rm out}}^{(2)}=
    -i\hat b_B(t_P+t_G),
    \label{xiaxibout}
\end{align}
where we define the final output of the mechanical modes as $\hat b_{j,{\rm out}}=\hat b_j(2t_P+t_G)$. Also, we define the normalized temporal modes of the second pulse as
\begin{align}
    \hat \xi_{j,{\rm in}}^{(2)}&=
    \sqrt{\frac{2G_j}{e^{2G_jt_P}-1}}
    \int_{t_P+t_G}^{2t_P+t_G}dse^{G_j(s-t_P-t_G)}\hat a_{j,{\rm in}}(s)\\
    \hat \xi_{j,{\rm out}}^{(2)}&=
    \sqrt{\frac{2G_j}{1-e^{-2G_jt_P}}}
    \int_{t_P+t_G}^{2t_P+t_G}dse^{-G_j(s-t_P-t_G)}\hat a_{j,{\rm out}}(s),
\end{align}
where $[\hat \xi_{j,{\rm in}}^{(2)},\hat \xi_{j,{\rm in}}^{(2)\dagger}]=[\hat \xi_{j,{\rm out}}^{(2)},\hat \xi_{j,{\rm out}}^{(2)\dagger}]=1$. Solving these equations combined with the solutions during the gravitational interaction, we obtain the output operators as
\begin{align}
    \label{xiathermal}
    &\hat \xi_{A,{\rm out}}^{(2)}=
    -C_{11}(t_G)\hat \xi_{A,{\rm in}}^{(1)}+C_{12}(t_G)\hat \xi_{B,{\rm in}}^{(1)}+in_A\\
    &\hat \xi_{B,{\rm out}}^{(2)}=
    C_{21}(t_G)\hat \xi_{A,{\rm in}}^{(1)}-C_{22}(t_G)\hat \xi_{B,{\rm in}}^{(1)}-in_B.
    \label{xibthermal}
\end{align}

\section{Density matrix under thermal noise effects}
\label{apdx:thermal}
In this section, we show the derivation of the density matrix from the characteristic function under thermal noise effects. Here, we assume the initial density matrix as $\rho_{\rm in}=\rho_{\rm in}^m\otimes\rho_{\rm in}^{(1)}\otimes\rho_{\rm in}^{(2)}\otimes\rho_{\rm th}$ consisting of the mechanical modes, the first optical modes, the second optical modes, and the environment. The reduced density matrix of the output state of the second optical modes is given by
\begin{align}
    \rho_{{\rm out},{\rm th}}^{(2)}&=
    \int\frac{d^2\alpha_A}{\pi}\int\frac{d^2\alpha_B}{\pi}
    \chi_{\rm out}(\alpha_A,\alpha_B)\hat D_A^{(2)}(-\alpha_A)\hat D_B^{(2)}(-\alpha_B),
\end{align}
where the characteristic function of the output state $\chi_{\rm out}(\alpha_A,\alpha_B)$ is
\begin{align}
    \chi_{\rm out}(\alpha_A,\alpha_B)&=
    {\rm tr}\left[\rho_{\rm in}\hat D_A^{(2)}(\alpha_A)\hat D_B^{(2)}(\alpha_B)\right],
\end{align}
and the displacement operator of the second optical mode $\hat D_j^{(2)}(\alpha_j)$ is
\begin{align}
    \hat D_j^{(2)}(\alpha_j)&=
    \exp[\alpha_j^*\hat \xi_{j,{\rm out}}^{(2)}-\alpha_j\hat \xi_{j,{\rm out}}^{(2)\dagger}].
\end{align}
Using the output operators \eqref{xiathermal} and \eqref{xibthermal}, the displacement operators are written as
\begin{align}
    \hat D_A^{(2)}(\alpha_A)&=\hat D_A^{(1)}(-\alpha_AC_{11}^*(t_G))\hat D_B^{(1)}(\alpha_AC_{12}^*(t_G))
    e^{i\alpha_A^*n_A^{11}+i\alpha_An_A^{11\dagger}}
    e^{i\alpha_A^*n_B^{12}+i\alpha_An_B^{12\dagger}}\\
    \hat D_B^{(2)}(\alpha_B)&=\hat D_A^{(1)}(\alpha_BC_{21}^*(t_G))\hat D_B^{(1)}(-\alpha_BC_{22}^*(t_G))
    e^{-i\alpha_B^*n_A^{21}-i\alpha_Bn_A^{21\dagger}}
    e^{-i\alpha_B^*n_B^{22}-i\alpha_Bn_B^{22\dagger}},
\end{align}
where we define the displacement operator of the first optical mode $\hat D_j^{(1)}(\alpha_j)=e^{\alpha_j^*\hat \xi_{j,{\rm in}}^{(1)}-\alpha_j\hat \xi_{j,{\rm in}}^{(1)\dagger}}$ and the thermal noise
\begin{align}
    n_j^{nm}&=
    \int_{t_P}^{t_P+t_G}dsC_{nm}(t_P+t_G-s)f_{j,{\rm th}}(s).
\end{align}
Using the Baker–Campbell–Hausdorff formula, we have
\begin{align}
    \hat D_A^{(2)}(\alpha_A)\hat D_B^{(2)}(\alpha_B)&=
    \exp\left[-\frac{1}{2}\alpha_A\alpha_B^*(C_{11}^*C_{21}+C_{12}^*C_{22})+\frac{1}{2}\alpha_A^*\alpha_B(C_{11}C_{21}^*+C_{12}C_{22}^*)\right]\nonumber\\
    &\quad\times
    \hat D_A^{(1)}(-\alpha_AC_{11}^*+\alpha_BC_{21}^*)
    \hat D_B^{(1)}(\alpha_AC_{12}^*-\alpha_BC_{22}^*)
    \nonumber\\
    &\quad\times
    e^{i\alpha_A^*n_A^{11}+i\alpha_An_A^{11\dagger}}
    e^{i\alpha_A^*n_B^{12}+i\alpha_An_B^{12\dagger}}
    e^{-i\alpha_B^*n_A^{21}-i\alpha_Bn_A^{21\dagger}}
    e^{-i\alpha_B^*n_B^{22}-i\alpha_Bn_B^{22\dagger}}.
\end{align}

Hence, the characteristic function is obtained by
\begin{align}
    \chi_{\rm out}(\alpha_A,\alpha_B)&=
    \exp\left[-\frac{1}{2}\alpha_A\alpha_B^*(C_{11}^*C_{21}+C_{12}^*C_{22})+\frac{1}{2}\alpha_A^*\alpha_B(C_{11}C_{21}^*+C_{12}C_{22}^*)\right]\nonumber\\
    &\quad\times
    {\rm tr}[\rho_{\rm in}^{(1)}\hat D_A^{(1)}(-\alpha_AC_{11}^*+\alpha_BC_{21}^*)\hat D_B^{(1)}(\alpha_AC_{12}^*-\alpha_BC_{22}^*)]\nonumber\\
    &\quad\times
    {\rm tr}\left[\rho_{\rm th}e^{i\alpha_A^*n_A^{11}+i\alpha_An_A^{11\dagger}}
    e^{i\alpha_A^*n_B^{12}+i\alpha_An_B^{12\dagger}}
    e^{-i\alpha_B^*n_A^{21}-i\alpha_Bn_A^{21\dagger}}
    e^{-i\alpha_B^*n_B^{22}-i\alpha_Bn_B^{22\dagger}}\right].
\end{align}

Using Eq.~\eqref{cmat}, the characteristic function can be decomposed as the product of a  pure evolution part and a thermal part as
\begin{align}
    \chi_{\rm out}(\alpha_A,\alpha_B)&=
    \chi_{\rm in}\chi_{\rm th},
\end{align}
where each part is
\begin{align}
    \chi_{\rm in}&=
    {\rm tr}[\rho_{\rm in}^{(1)}\hat D_A^{(1)}(-\alpha_AC_{11}^*+\alpha_BC_{21}^*)\hat D_B^{(1)}(\alpha_AC_{12}^*-\alpha_BC_{22}^*)]\\
    \chi_{\rm th}&=
    {\rm tr}\left[\rho_{\rm th}e^{i\alpha_A^*n_A^{11}+i\alpha_An_A^{11\dagger}}
    e^{-i\alpha_B^*n_A^{21}-i\alpha_Bn_A^{21\dagger}}
    e^{i\alpha_A^*n_B^{12}+i\alpha_An_B^{12\dagger}}
    e^{-i\alpha_B^*n_B^{22}-i\alpha_Bn_B^{22\dagger}}\right]
\end{align}

Let us compute the thermal part $\chi_{\rm th}$. Since the thermal noise is Gaussian, the commutation relation of the thermal fluctuation is a c-number. Using Wick's theorem, we have
\begin{align}
    \label{apdx:braket}
    \chi_{\rm th}
    &=e^{\frac{1}{2}[i\alpha_A^*n_A^{11}+i\alpha_An_A^{11\dagger},-i\alpha_B^*n_A^{21}-i\alpha_Bn_A^{21\dagger}]}
    e^{\frac{1}{2}[i\alpha_A^*n_B^{12}+i\alpha_An_B^{12\dagger},-i\alpha_B^*n_B^{22}-i\alpha_Bn_B^{22\dagger}]}
    \nonumber\\
    &\quad\times
    \braket{e^{i\alpha_A^*n_A^{11}+i\alpha_An_A^{11\dagger}-i\alpha_B^*n_A^{21}-i\alpha_Bn_A^{21\dagger}}}
    \braket{e^{i\alpha_A^*n_B^{12}+i\alpha_An_B^{12\dagger}-i\alpha_B^*n_B^{22}-i\alpha_Bn_B^{22\dagger}}}
    \nonumber\\
    &=e^{\frac{1}{2}[i\alpha_A^*n_A^{11}+i\alpha_An_A^{11\dagger},-i\alpha_B^*n_A^{21}-i\alpha_Bn_A^{21\dagger}]}
    e^{\frac{1}{2}[i\alpha_A^*n_B^{12}+i\alpha_An_B^{12\dagger},-i\alpha_B^*n_B^{22}-i\alpha_Bn_B^{22\dagger}]}
    \nonumber\\
    &\quad\times
    e^{-\frac{1}{2}\braket{(\alpha_A^*n_A^{11}+\alpha_An_A^{11\dagger}-\alpha_B^*n_A^{21}-\alpha_Bn_A^{21\dagger})^2}}
    e^{-\frac{1}{2}\braket{(\alpha_A^*n_B^{12}+\alpha_An_B^{12\dagger}-\alpha_B^*n_B^{22}-\alpha_Bn_B^{22\dagger})^2}}.
\end{align}
The correlation of the thermal fluctuation is given by $\braket{\{f_{j,{\rm th}}(t),f_{j,{\rm th}}(t^\prime)^\dagger\}}/2=2\gamma_m N_{{\rm th}}\delta(t-t^\prime)$. Then the correlation of $n_j^{k\ell}$ is
\begin{align}
    \braket{\{n_j^{k\ell},n_{j}^{k^\prime \ell^\prime\dagger}\}}/2&=
    \int_{t_P}^{t_P+t_G}ds\int_{t_P}^{t_P+t_G}ds^\prime
    C_{k\ell}(t_P+t_G-s)C_{k^\prime \ell^\prime}^*(t_P+t_G-s^\prime)
    \braket{\{f_{j,{\rm th}}(s),f_{j,{\rm th}}^\dagger(s^\prime)\}}/2
    \nonumber\\
    &=
    2\gamma_m N_{{\rm th}}
    \int_{t_P}^{t_P+t_G}ds\int_{t_P}^{t_P+t_G}ds^\prime
    C_{k\ell}(t_P+t_G-s)C_{k^\prime \ell^\prime}^*(t_P+t_G-s^\prime)\delta(s-s^\prime)
    \nonumber\\
    &=
    2\gamma_m N_{{\rm th}}
    \int_{t_P}^{t_P+t_G}ds
    C_{k\ell}(t_P+t_G-s)C_{k^\prime \ell^\prime}^*(t_P+t_G-s).
\end{align}
Assuming the commutation relation of the thermal noise $[f_{j,{\rm th}}(t),f_{j,{\rm th}}^\dagger(t^\prime)]=2\gamma_m \delta(t-t^\prime)$, we have
\begin{align}
    [n_j^{k\ell},n_{j}^{k^\prime \ell^\prime\dagger}]&=
    \int_{t_P}^{t_P+t_G}ds\int_{t_P}^{t_P+t_G}ds^\prime
    C_{k\ell}(t_P+t_G-s)C_{k^\prime \ell^\prime}^*(t_P+t_G-s^\prime)
    [f_{j,{\rm th}}(s),f_{j,{\rm th}}^\dagger(s^\prime)]
    \nonumber\\
    &=
    2\gamma_m\int_{t_P}^{t_P+t_G}ds
    C_{k\ell}(t_P+t_G-s)C_{k^\prime \ell^\prime}^*(t_P+t_G-s).
\end{align}
Hence, each part of the right hand side of Eq.~\eqref{apdx:braket} is
\begin{align}
    e^{\frac{1}{2}[i\alpha_A^*n_A^{11}+i\alpha_An_A^{11\dagger},-i\alpha_B^*n_A^{21}-i\alpha_Bn_A^{21\dagger}]}
    &=
    e^{\frac{1}{2}
    \alpha_A^*\alpha_B[n_A^{11},n_A^{21\dagger}]
    +\frac{1}{2}\alpha_A\alpha_B^*[n_A^{11\dagger},n_A^{21}]}
    =e^{{\rm Re}[\alpha_A^*\alpha_B[n_A^{11},n_A^{21\dagger}]]}
    \nonumber\\
    e^{\frac{1}{2}[i\alpha_A^*n_B^{12}+i\alpha_An_B^{12\dagger},-i\alpha_B^*n_B^{22}-i\alpha_Bn_B^{22\dagger}]}
    &=
    e^{\frac{1}{2}
    \alpha_A^*\alpha_B[n_B^{12},n_B^{22\dagger}]
    +\frac{1}{2}\alpha_A\alpha_B^*[n_B^{12\dagger},n_B^{22}]}
    =e^{{\rm Re}[\alpha_A^*\alpha_B[n_B^{12},n_B^{22\dagger}]]}
    \nonumber\\
    e^{-\frac{1}{2}\braket{(\alpha_A^*n_A^{11}+\alpha_An_A^{11\dagger}-\alpha_B^*n_A^{21}-\alpha_Bn_A^{21\dagger})^2}}
    &=e^{-\frac{1}{2}|\alpha_A|^2\braket{\{n_A^{11},n_A^{11\dagger}\}}
    -\frac{1}{2}\alpha_A^*\alpha_B\braket{\{n_A^{11},n_A^{21\dagger}\}}
    -\frac{1}{2}\alpha_A\alpha_B^*\braket{\{n_A^{11\dagger},n_A^{21}\}}
    +\frac{1}{2}|\alpha_B|^2\braket{\{n_A^{21},n_A^{21\dagger}\}}}
    \nonumber\\
    e^{-\frac{1}{2}\braket{(\alpha_A^*n_B^{12}+\alpha_An_B^{12\dagger}-\alpha_B^*n_B^{22}-\alpha_Bn_B^{22\dagger})^2}}
    &=e^{-\frac{1}{2}|\alpha_A|^2\braket{\{n_B^{12},n_B^{12\dagger}\}}
    -\frac{1}{2}\alpha_A^*\alpha_B\braket{\{n_B^{12},n_B^{22\dagger}\}}
    -\frac{1}{2}\alpha_A\alpha_B^*\braket{\{n_B^{12^\dagger},n_B^{22}\}}
    +\frac{1}{2}|\alpha_B|^2\braket{\{n_B^{22},n_B^{22\dagger}\}}}
    \nonumber
\end{align}
Calculating these terms, we derive
\begin{align}
    \chi_{\rm th}&=
    e^{-(1-e^{-2\gamma_m t_G})N_{\rm th}(|\alpha_A|^2+|\alpha_B|^2)}.
\end{align}

Next, we consider the characteristic function of pure evolution part. Here, we assume the initial $n$-photon state of the first optical mode A,
\begin{align}
    \rho_{{\rm in},n}^{(1)}=\ket{n,0}\bra{n,0}.
\end{align}
Then, the characteristic function is obtained by
\begin{align}
    \chi_{{\rm in},n}&=
    {\rm tr}[\ket{n}\bra{n}\hat D_A^{(1)}(-\alpha_AC_{11}^*+\alpha_BC_{21}^*)]{\rm tr}[\ket{0}\bra{0}\hat D_B^{(1)}(\alpha_AC_{12}^*-\alpha_BC_{22}^*)]\nonumber\\
    &=e^{-(|\alpha_A|^2+|\alpha_B|^2)/2}
    L_n(|\alpha_AC_{11}^*-\alpha_BC_{21}^*|^2),
\end{align}
where the Laguerre polynomial $L_n(x)$ is defined by
\begin{align}
    L_n(x)&=
    \sum_{k=0}^n\binom{n}{k}\frac{(-x)^k}{k!}.
\end{align}
The total characteristic function of the output state is
\begin{align}
    \chi_{{\rm out},n}(\alpha_A,\alpha_B)&=
    e^{-(1+2(1-e^{-2\gamma_m t_G})N_{\rm th})(|\alpha_A|^2+|\alpha_B|^2)/2}
    L_n(|\alpha_AC_{11}^*-\alpha_BC_{21}^*|^2),
\end{align}
and the reduced density matrix is
\begin{align}
    \rho_{{\rm out},n,{\rm th}}^{(2)}&=
    \int\frac{d^2\alpha_A}{\pi}\int\frac{d^2\alpha_B}{\pi}
    e^{-(1+2(1-e^{-2\gamma_m t_G})N_{\rm th})(|\alpha_A|^2+|\alpha_B|^2)/2}
    L_n(|\alpha_AC_{11}^*-\alpha_BC_{21}^*|^2)
    \hat D_A^{(2)}(-\alpha_A)\hat D_B^{(2)}(-\alpha_B).
\end{align}
The displacement operator in the Fock basis is written as
\begin{align}
    \bra{m}\hat D_j(\alpha_j)\ket{\ell}&=
    e^{-|\alpha_j|^2/2}\sqrt{\frac{{\rm min}(m,\ell)!}{{\rm max}(m,\ell)!}}
    \left\{ \begin{array}{cc}
    (\alpha_j^*)^{\ell-m}L_m^{(\ell-m)}(|\alpha_j|^2), & \ell\ge m \\
    (-\alpha_j)^{m-\ell}L_\ell^{(m-\ell)}(|\alpha_j|^2), & m>\ell
    \end{array} \right.
\end{align}
where $L_m^{(\ell-m)}(x)$ is the associated Laguerre polynomial
\begin{align}
    L_m^{(\ell-m)}(x)&=
    \sum_{k=0}^m\binom{\ell}{m-k}\frac{(-x)^k}{k!}.
\end{align}
Using $C_{11}^*=C_{11}$ and $C_{21}^*=-C_{21}$, we can write the Laguerre polynomial
\begin{align}
    L_n(|\alpha_AC_{11}^*-\alpha_BC_{21}^*|^2)&=
    \sum_{k=0}^n\binom{n}{k}
    \frac{(-1)^k}{k!}(|\alpha_AC_{11}+\alpha_BC_{21}|^2)^k\nonumber\\
    &=\sum_{k=0}^n\binom{n}{k}
    \frac{(-1)^k}{k!}(\alpha_AC_{11}+\alpha_BC_{21})^k(\alpha_A^*C_{11}-\alpha_B^*C_{21})^k\nonumber\\
    &=\sum_{k=0}^n\binom{n}{k}
    \frac{1}{k!}
    \sum_{k^\prime=0}^k\binom{k}{k^\prime}
    \sum_{k^{\prime\prime}=0}^k\binom{k}{k^{\prime\prime}}
    (-1)^{k^{\prime\prime}}
    (C_{11})^{k^\prime+k^{\prime\prime}}
    (C_{21})^{2k-k^\prime-k^{\prime\prime}}
    \nonumber\\
    &\quad\times
    (\alpha_A)^{k^\prime}(\alpha_A^*)^{k^{\prime\prime}}
    (\alpha_B)^{k-k^\prime}
    (\alpha_B^*)^{k-k^{\prime\prime}}\nonumber
\end{align}
Hence, the component of the reduced density matrix in the Fock basis $(\rho_{{\rm out},n,{\rm th}}^{(2)})_{mm^\prime,\ell\ell^\prime}=\bra{m,m^\prime}\rho_{{\rm out},n,{\rm th}}^{(2)}\ket{\ell,\ell^\prime}$ is
\begin{align}
    (\rho_{{\rm out},n,{\rm th}}^{(2)})_{mm^\prime,\ell\ell^\prime}&=
    \sqrt{\frac{{\rm min}(m,\ell)!{\rm min}(m^\prime,\ell^\prime)!}{{\rm max}(m,\ell)!{\rm max}(m^\prime,\ell^\prime)!}}
    \sum_{k=0}^n
    \sum_{k^\prime=0}^k
    \sum_{k^{\prime\prime}=0}^k
    J_{k,k^\prime,k^{\prime\prime}}
    I_A^{k,k^\prime,k^{\prime\prime}}I_B^{k,k^\prime,k^{\prime\prime}},
    \label{densitythermal}
\end{align}
where
\begin{align}
    &J_{k,k^\prime,k^{\prime\prime}}=
    \frac{1}{k!}
    \binom{n}{k}
    \binom{k}{k^\prime}
    \binom{k}{k^{\prime\prime}}
    (-1)^{k^{\prime\prime}}
    (C_{11})^{k^\prime+k^{\prime\prime}}
    (C_{21})^{2k-k^\prime-k^{\prime\prime}}
    \\
    &I_A^{k,k^\prime,k^{\prime\prime}}=
    \int\frac{d^2\alpha_A}{\pi}
    e^{-(1+(1-e^{-2\gamma_m t_G})N_{\rm th})|\alpha_A|^2}
    \left\{\begin{array}{cc}
    (\alpha_A)^{k^{\prime}}(\alpha_A^*)^{\ell-m+k^{\prime\prime}}L_m^{(\ell-m)}(|\alpha_A|^2), &  \ell\ge m\\
    (-1)^{m-\ell}(\alpha_A)^{m-\ell+k^\prime}(\alpha_A^*)^{k^{\prime\prime}}
    L_\ell^{(m-\ell)}(|\alpha_A|^2), & m>\ell 
    \end{array}\right.
    \\
    &I_B^{k,k^\prime,k^{\prime\prime}}=
    \int\frac{d^2\alpha_B}{\pi}
    e^{-(1+(1-e^{-2\gamma_m t_G})N_{\rm th})|\alpha_B|^2}
    \left\{\begin{array}{cc}
    (\alpha_B)^{k-k^{\prime}}(\alpha_B^*)^{\ell^\prime-m^\prime+k-k^{\prime\prime}}L_{m^\prime}^{(\ell^\prime-m^\prime)}(|\alpha_B|^2), &  \ell^\prime\ge m^\prime\\
    (-1)^{m^\prime-\ell^\prime}(\alpha_B)^{m^\prime-\ell^\prime+k-k^\prime}(\alpha_B^*)^{k-k^{\prime\prime}}L_{\ell^\prime}^{(m^\prime-\ell^\prime)}(|\alpha_B|^2), & m^\prime>\ell^\prime 
    \end{array}\right.
\end{align}
Introducing the variables $\alpha_j=r_je^{i\theta_j}$ and $\alpha_j^*=r_je^{-i\theta_j}$, we can derive
\begin{align}
    I_A^{k,k^\prime,k^{\prime\prime}}&=
    \int_0^\infty dr_Ar_A\int_0^{2\pi}d\theta_A\frac{1}{\pi}
    e^{-(1+(1-e^{-2\gamma_m t_G})N_{\rm th})r_A^2}
    \left\{\begin{array}{cc}
    \sum_{u=0}^m\binom{\ell}{m-u}\frac{(-1)^u}{u!}
    r_A^{\ell-m+k^\prime+k^{\prime\prime}+2u}
    e^{i\theta_A(k^\prime-\ell+m-k^{\prime\prime})}, &  \ell\ge m\\
    \sum_{u=0}^\ell\binom{m}{\ell-u}\frac{(-1)^u}{u!}(-1)^{m-\ell}
    r_A^{m-\ell+k^\prime+k^{\prime\prime}+2u}
    e^{i\theta_A(m-\ell+k^\prime-k^{\prime\prime})}, & m>\ell 
    \end{array}\right.
    \nonumber\\
    &=2\int_0^\infty dr_A
    e^{-(1+(1-e^{-2\gamma_m t_G})N_{\rm th})r_A^2}
    \delta_{k^{\prime\prime},k^\prime-\ell+m}
    \left\{\begin{array}{cc}
    \sum_{u=0}^m\binom{\ell}{m-u}\frac{(-1)^u}{u!}
    r_A^{2k^\prime+2u+1}, &  \ell\ge m\\
    \sum_{u=0}^\ell\binom{m}{\ell-u}\frac{(-1)^u}{u!}(-1)^{m-\ell}
    r_A^{2m-2\ell+2k^\prime+2u+1}, & m>\ell 
    \end{array}\right.
    \nonumber\\
    &=
    \delta_{k^{\prime\prime},k^\prime-\ell+m}
    \left\{\begin{array}{cc}
    \sum_{u=0}^m\binom{\ell}{m-u}\frac{(-1)^u}{u!}
    \frac{\Gamma\left(k^\prime+u+1\right)}{(1+(1-e^{-2\gamma_m t_G})N_{\rm th})^{k^\prime+u+1}}, &  \ell\ge m\\
    \sum_{u=0}^\ell\binom{m}{\ell-u}\frac{(-1)^u}{u!}(-1)^{m-\ell}
    \frac{\Gamma\left(m-\ell+k^\prime+u+1\right)}{(1+(1-e^{-2\gamma_m t_G})N_{\rm th})^{m-\ell+k^\prime+u+1}}, & m>\ell 
    \end{array}\right.
\end{align}
and
\begin{align}
    I_B^{k,k^\prime,k^{\prime\prime}}&=
    \delta_{k^{\prime\prime},k^\prime+\ell^\prime-m^\prime}
    \left\{\begin{array}{cc}
    \sum_{v=0}^{m^\prime}\binom{\ell^\prime}{m^\prime-v}\frac{(-1)^v}{v!}
    \frac{\Gamma\left(k-k^\prime+v+1\right)}{(1+(1-e^{-2\gamma_m t_G})N_{\rm th})^{k-k^\prime+v+1}}, &  \ell^\prime\ge m^\prime\\
    \sum_{v=0}^{\ell^\prime}\binom{m^\prime}{\ell^\prime-v}\frac{(-1)^v}{v!}(-1)^{m^\prime-\ell^\prime}
    \frac{\Gamma\left(m^\prime-\ell^\prime+k-k^\prime+v+1\right)}{(1+(1-e^{-2\gamma_m t_G})N_{\rm th})^{m^\prime-\ell^\prime+k-k^\prime+v+1}}, & m^\prime>\ell^\prime 
    \end{array}\right.
\end{align}
where we used the following relation
\begin{align}
    \int_0^\infty dr e^{-ar^2}r^n&=
    \frac{1}{2}a^{-(n+1)/2}\Gamma\left(\frac{n+1}{2}\right)
\end{align}
with a Gamma function $\Gamma((n+1)/2)$ for $a>0$ and $n>-1$. The reduced density matrix is derived by substituting the results of $J_{k,k^\prime,k^{\prime\prime}}$, $I_A^{k,k^\prime,k^{\prime\prime}}$, and $I_B^{k,k^\prime,k^{\prime\prime}}$ into Eq.~\eqref{densitythermal}.

\section{Separability-preservation condition for finite gravitational interaction time}
\label{apdx: necessary}
\subsection{Lossless case}
Here, we derive the separability-preservation condition without expanding in the gravitational interaction angle $\theta\equiv g_Gt_G$. We first consider the lossless case. As discussed in Sec.~\ref{sec:universality}, the PPT condition for the output covariance matrix can be written as $\tilde{\bm V} = \bm V_{\rm in} +\bm\Sigma_{\rm PT} +\bm\Omega \ge0$, where $\bm\Omega$ is defined in Eq.~\eqref{Omegamat}. Since any initially
separable two-mode Gaussian state satisfies $\bm V_{\rm in}+\bm\Sigma_{\rm PT}\ge0$, it immediately follows that $\tilde{\bm V}\ge0$ for all such input states whenever $\bm\Omega\ge0$. Thus, $\bm\Omega\ge0$ is a sufficient condition for separability preservation.

The matrix $\bm\Omega$ has eigenvalues $\lambda_\pm(\bm \Omega)=\Omega_1\pm\sqrt{\Omega_2^2+\Omega_3^2}$. Therefore, $\bm\Omega\ge0$ is equivalent to $\Omega_1\ge\sqrt{\Omega_2^2+\Omega_3^2}$. Expanding only in $\gamma_mt_G\ll1$, we obtain
\begin{align}
    \lambda_{\rm min}(\bm\Omega)
    \simeq
    -2|\sin[g_Gt_G]|
    +4\gamma_mt_GN_{\rm th}.
\end{align}
Thus, to this order, the sufficient condition for separability preservation is
\begin{align}
    2\gamma_mt_GN_{\rm th}
    \ge
    |\sin(g_Gt_G)|.
    \label{ap lossysep}
\end{align}
This reduces to $2\gamma_mN_{\rm th}\ge g_G$ when the additional short-interaction approximation $g_Gt_G\ll1$ is used.

We next discuss when this sufficient condition is also necessary. We do not prove the converse for arbitrary finite values of $g_Gt_G$. Instead, we show that the converse holds in the parameter regime where $\Omega_2\ge0$, which includes the short-interaction regime $g_Gt_G\ll1$ used in the main text. To this end, we prove the contrapositive within this regime: if $\bm\Omega\not\ge0$, then there exists an initially separable two-mode Gaussian state for which $\tilde{\bm V}\not\ge0$. This means that the evolution can generate entanglement from an initially separable Gaussian state.

We choose the initially separable squeezed product state $\bm V_{\rm in} = {\rm diag}\{e^{2\zeta},e^{-2\zeta},e^{2\zeta},e^{-2\zeta}\}$. This state is sufficient for proving the converse direction in the parameter regime considered below. Introducing the complex vector $\bm x = \alpha(1,ie^{2\zeta},0,0)^{\rm T} + \beta(0,0,1,-ie^{2\zeta})^{\rm T}$ with arbitrary complex numbers $\alpha$ and $\beta$, one finds $\bm x^\dagger \left( \bm V_{\rm in}+\bm\Sigma_{\rm PT} \right) \bm x = 0$ so that $\bm x^\dagger\tilde{\bm V}\bm x = \bm x^\dagger\bm\Omega\bm x$. The quadratic form on the right-hand side can be written as $\bm x^\dagger\bm\Omega\bm x = \bm z^\dagger\tilde{\bm\Omega}\bm z$, where $\bm z=(\alpha,\beta)^{\rm T}$ and
\begin{align}
    \tilde{\bm\Omega}
    =
    \left(
    \begin{array}{cc}
        \tilde\Omega_1 & i\tilde\Omega_2\\
        -i\tilde\Omega_2 & \tilde\Omega_1
    \end{array}
    \right),
\end{align}
with $\tilde \Omega_1=\Omega_1(1+e^{4\zeta})-2\Omega_2e^{2\zeta}$ and $\tilde \Omega_2=\Omega_3(1-e^{4\zeta})$. The two eigenvalues of $\tilde{\bm\Omega}$ are
\begin{align}
    \lambda_\pm(\tilde{\bm\Omega})
    =
    (\Omega_1\pm\Omega_3)e^{4\zeta}
    -2\Omega_2e^{2\zeta}
    +\Omega_1\mp\Omega_3.
\end{align}
From the definitions of the components of $\bm\Omega$, we have $\Omega_1\ge0$ and $\Omega_2,\Omega_3\in\mathbb{R}$. In the following, we first assume $\Omega_2>0$. Under the condition $\lambda_{\rm min}(\bm \Omega)<0$, the minimum eigenvalue of $\tilde{\bm\Omega}$ is then obtained as
\begin{align}
    \lambda_{\rm min}(\tilde{\bm\Omega})
    =
    \left\{
    \begin{array}{lll}
        -2(\Omega_2e^{2\zeta}-\Omega_1)\to -\infty
        & {\rm for}~\Omega_1=|\Omega_3|
        & (\zeta\to\infty),
        \\[1ex]
        \Omega_1-|\Omega_3|<0
        & {\rm for}~\Omega_1<|\Omega_3|
        & (\zeta\to\infty),
        \\[1ex]
        \displaystyle
        \frac{\Omega_1^2-\Omega_2^2-\Omega_3^2}
        {\Omega_1-|\Omega_3|}<0
        & {\rm for}~\Omega_1>|\Omega_3|
        & \left(e^{2\zeta}=
        \frac{\Omega_2}{\Omega_1-|\Omega_3|}\right).
    \end{array}
    \right.
\end{align}
Therefore, whenever $\lambda_{\rm min}(\bm\Omega)<0$ and $\Omega_2>0$, one can choose the squeezing parameter $\zeta$ such that $\lambda_{\rm min}(\tilde{\bm\Omega})<0$. The boundary case $\Omega_2=0$ can be treated separately. In this case, the condition $\lambda_{\rm min}(\bm\Omega)<0$ reduces to $\Omega_1<|\Omega_3|$. This is exactly the second case above, for which $\lambda_{\rm min}(\tilde{\bm\Omega})<0$ in the large squeezing limit. If instead $\Omega_1\ge|\Omega_3|$, then $\lambda_{\rm min}(\bm\Omega)\ge0$, so this case is outside the regime $\bm\Omega\not\ge0$. Hence the converse proof also holds when $\Omega_2=0$.

Therefore, under the assumption $\Omega_2\ge0$, whenever $\lambda_{\rm min}(\bm\Omega)<0$, one can optimize the squeezing parameter $\zeta$ so that $\lambda_{\rm min}(\tilde{\bm\Omega})<0$. This implies that there exists an initially separable squeezed product state for which $\tilde{\bm V}\not\ge0$ and the condition $\bm\Omega\ge0$ is not only sufficient but also necessary for separability preservation.

For finite $g_Gt_G$, however, $\Omega_2$ is not necessarily positive. Therefore, outside the regime $\Omega_2\ge0$, the condition $\bm\Omega\ge0$ should be regarded as a sufficient condition within the present proof. In particular, in the short-interaction regime used in the main text, $g_Gt_G\ll1$, one has $\Omega_2\simeq0$ to the first order, and the separability-preservation condition becomes necessary and sufficient. In this limit, the leading-order condition reduces to $2\gamma_mN_{\rm th}\ge g_G$.

\subsection{Lossy case}
\label{apdx: lossy universal}
We consider the same system as in Sec.~\ref{sec:universality}, but now apply the loss channel to the final time evolved covariance matrix. The resulting post-loss covariance matrix is
\begin{align}
    \bm{V}_{\text{post-loss}}=
    \eta\left(\bm M\bm V_{\rm in}\bm M^{\rm T}+\bm V_{\rm th}\right)+(1-\eta)\bm 1,
\end{align}
where we assume $\eta_{A} = \eta_{B} \equiv \eta$ and $\bar{n} = 0$.
We first evaluate the entanglement negativity for the initially squeezed product state $\bm V_{\rm in}={\rm diag}\{e^{2\zeta},e^{-2\zeta},e^{2\zeta},e^{-2\zeta}\}$. Assuming $\gamma_m t_G\ll1$, we obtain the entanglement negativity as
\begin{align}
    \mathcal N_{G,{\rm loss}}^{\rm sq}&=\frac{1}{2(\sqrt{f_{\rm loss}}-\eta|\sinh[2\zeta]\sin[2g_Gt_G]|)}-\frac{1}{2},
\end{align}
where
\begin{align}
    f_{\rm loss}&=1+2\eta(1-\eta)(\cosh[2\zeta]-1)+8\gamma_mt_GN_{\rm th}\eta(1+\eta(\cosh[2\zeta]-1))+16\eta^2(\gamma_mt_GN_{\rm th})^2+\eta^2\sinh^2[2\zeta]\sin^2[2g_Gt_G]
\end{align}
The condition $\mathcal N_{G,{\rm loss}}^{\rm sq}>0$ is equivalent to
\begin{align}
    4\eta\gamma_mt_GN_{\rm th}<
    \sqrt{1+\eta^2\sinh^2[2\zeta]+2\eta|\sinh[2\zeta]\sin[2g_Gt_G]|}-(1-\eta+\eta\cosh[2\zeta]).
    \label{lossyneg}
\end{align}
This inequality reduces to Eq.~\eqref{ngsq_condition} in the lossless limit $\eta=1$.

Next, we derive a universal bound on entanglement generation that is independent of the initial state. According to the positive partial transpose condition, the state is separable if and only if the following inequality holds for the two-mode Gaussian state:
\begin{equation}
    \bm{V}_{\text{post-loss}} + 
    \bm \Sigma_{\rm PT}\geq 0.
\end{equation}
Moving the time evolution piece away from $\bm{V}_{\rm in}$, the relevant inequality becomes
\begin{equation} 
    \bm V_{\rm in}+\frac{1}{\eta}\bm M^{-1}\bm \Sigma_{\rm PT}(\bm M^{\rm T})^{-1}+\bm M^{-1}\bm V_{\rm th}(\bm M^{\rm T})^{-1}+\frac{1-\eta}{\eta}e^{2\gamma_{m}t_{G}}\bm 1\geq 0.
\end{equation}
As in the lossless case, we can rewrite the left-hand side as
\begin{equation}
    \widetilde{\bm{V}}_{\text{post-loss}} =\bm{V}_{\text{in}}+\bm{\Sigma}_{\text{PT}}+\bm{\Omega}_{\rm loss}
\end{equation}
where
\begin{equation}
    \begin{split}
        \bm{\Omega}_{\rm loss} &= 
        \begin{pmatrix}
            \Omega_{1,{\rm loss}} & i\Omega_{2,{\rm loss}} & i\Omega_{3,{\rm loss}} & 0 \\
            -i\Omega_{2,{\rm loss}} & \Omega_{1,{\rm loss}} & 0 & i\Omega_{3,{\rm loss}} \\
            -i\Omega_{3,{\rm loss}} & 0 & \Omega_{1,{\rm loss}} & -i\Omega_{2,{\rm loss}} \\
            0 & -i\Omega_{3,{\rm loss}} & i\Omega_{2,{\rm loss}} & \Omega_{1,{\rm loss}}
        \end{pmatrix}\\
        \Omega_{1,{\rm loss}}&=-2 N_{\mathrm{th}} + e^{2 t_G \gamma_m}\!\left(-1 + 2 N_{\mathrm{th}} + \frac{1}{\eta}\right)\\
        \Omega_{2,{\rm loss}}&=-1+ \frac{e^{2 t_G \gamma_m}\cos[2 g_G t_G]}{\eta},\quad
        \Omega_{3,{\rm loss}}=\frac{e^{2 t_G \gamma_m}\sin[2 g_G t_G]}{\eta}
    \end{split}
\end{equation}
The smallest eigenvalue of $\bm{\Omega}_{\rm loss}$ is
\begin{equation}
\begin{split}
    &\lambda_{\text{min}}(\bm{\Omega}_{\rm loss}) = \Omega_{1,{\rm loss}}-\sqrt{\Omega_{2,{\rm loss}}^2+\Omega_{3,{\rm loss}}^2}.
\end{split}
\end{equation}
The claim is identical to Sec.~\ref{sec:universality}: an initial separable two-mode Gaussian state remains separable under the RWA with loss if $\bm{\Omega}_{\rm loss} \geq 0$. The forward implication is trivial: if $\bm{\Omega}_{\rm loss} \geq 0$, then $\widetilde{\bm{V}}_{\text{post-loss}}  \geq 0$ for any initial separable two-mode Gaussian state. Since the form of $\bm \Omega_{\rm loss}$ is the same as that of the lossless $\bm \Omega$, the properties of its components also carry over: $\Omega_{1,{\rm loss}}\ge0$ and $\Omega_{2,{\rm loss}},\Omega_{3,{\rm loss}}\in\mathbb{R}$. Hence, assuming $\Omega_{2,{\rm loss}}\ge0$, the condition $\bm \Omega_{\rm loss}\ge0$ is necessary and sufficient for preserving the separability.

Assuming $\gamma_mt_G\ll1$, the smallest eigenvalue of $\bm\Omega_{\rm loss}$ becomes
\begin{align}
    \lambda_{\rm min}(\bm\Omega_{\rm loss})
    &\simeq
    4\gamma_mt_GN_{\rm th}-\frac{\sqrt{(1-\eta)^2+4\eta\sin^2[g_Gt_G]}-(1-\eta)}{\eta},
    \label{eq:lambdaloss_gamma_small}
\end{align}
where we use the high-temperature regime $N_{\rm th}\gg1$. 
Therefore, the condition $\lambda_{\rm min}(\bm\Omega_{\rm loss})\ge0$
implies
\begin{align}
    4\eta\gamma_m t_G N_{\rm th}
    \ge
    \sqrt{(1-\eta)^2+4\eta\sin^2[g_Gt_G]}-(1-\eta).
    \label{eq:lossysep condition}
\end{align}
In the limit $\eta\to1$, this reduces to the lossless condition for the separability preservation, Eq.~\eqref{ap lossysep}. If we further assume $g_Gt_G\ll1$, then $\Omega_{2,{\rm loss}}\simeq(-1+1/\eta)+2\gamma_mt_G/\eta\ge0$. In this regime, the condition in Eq.~\eqref{eq:lossysep condition} is therefore both necessary and sufficient for separability preservation. For
finite $g_Gt_G$, however, $\Omega_{2,{\rm loss}}\simeq-1+\cos[2g_Gt_G](1+2\gamma_mt_G)/\eta$ is not necessarily positive, so Eq.~\eqref{eq:lossysep condition} should be regarded as a sufficient condition unless the positivity of $\Omega_{2,{\rm loss}}$ is separately guaranteed.

It is also useful to compare this condition with the explicit negativity condition obtained for the initial product squeezed state. In the regime $g_Gt_G\ll1$, the right-hand side of Eq.~\eqref{lossyneg} is maximized by choosing the squeezing parameter such that
\begin{align}
    |\tanh[2\zeta]|
    \simeq
    \frac{2|g_Gt_G|}
    {\sqrt{(1-\eta)^2+4\eta(g_Gt_G)^2}} .
    \label{eq:zeta_opt_loss}
\end{align}
With this choice, the optimized condition obtained from Eq.~\eqref{lossyneg} becomes
\begin{align}
    4\eta\gamma_m t_GN_{\rm th}
    \ge
    \sqrt{(1-\eta)^2+4\eta(g_Gt_G)^2}-(1-\eta).
\end{align}
This agrees with Eq.~\eqref{eq:lossysep condition} in the same limit $\sin[g_Gt_G]\simeq g_Gt_G$. Thus, in the regime $g_Gt_G\ll1$, the bound for the lossy case derived from $\bm\Omega_{\rm loss}$ is saturated by an appropriately chosen product squeezed input state. For finite measurement loss satisfying $4(g_Gt_G)^2<1-\eta$, this optimal squeezing is finite. As the loss vanishes, $\eta\to1$, Eq.~\eqref{eq:zeta_opt_loss} gives $\tanh[2\zeta]\to1$, i.e., $\zeta\to\infty$, recovering the lossless result that arbitrarily large squeezing approaches the threshold.

\section{Derivation of Post-Loss Covariance Matrix}
\label{apdx: post loss CM}
Suppose that a loss channel acts on two subsystems. We collect the quadrature operators of the system (s) and environment (e) into the column vector $\bm r_{\text{tot}} = (\bm r_{s},\bm r_{e})^{\rm T}$, where $\bm r_{s}\in\mathbb{R}^{2n}$ and $\bm r_{e}\in\mathbb{R}^{2m}$. Under a symplectic transformation $\bm S$, the first moment $\bar{\bm r}$ transforms as $\bm S\bar{\bm r}$ and the covariance matrix $\bm V$ transforms as $\bm S\bm V \bm S^{\rm T}$. We partition $\bm S$ into system and environment blocks as
\begin{equation}
    \bm S = \begin{pmatrix}
        \bm X & \bm Z\\
        \bm W & \bm T
    \end{pmatrix}
\end{equation}
so that
\begin{equation}
    \begin{pmatrix}
        \bm r'_{s}\\
        \bm r'_{e}
    \end{pmatrix}
    =
    \begin{pmatrix}
        \bm X & \bm Z\\
        \bm W & \bm T
    \end{pmatrix}
    \begin{pmatrix}
        \bm r_{s}\\
        \bm r_{e}
    \end{pmatrix}
\end{equation}
The first moment of the system then transforms as
\begin{equation}
    \bar{\bm r}_{s}' = \bm X\bar{\bm r}_{s}+\bm Z\bar{\bm r}_{e}
\end{equation}
Assume the initial total covariance matrix is
\begin{equation}
    \bm V_{\text{tot}} = \begin{pmatrix}
        \bm V_{s} & \bm V_{\rm int}\\
        \bm V_{\rm int}^{\rm T} & \bm V_{e}
    \end{pmatrix}
\end{equation}
where we set $\bm V_{\rm int} = 0$, because we assume that the system and environment are initially uncorrelated. Evaluating $\bm V'_{\text{tot}} = \bm S\bm V_{\text{tot}}\bm S^{\rm T}$, the system block (top left) becomes 
\begin{equation}
    \bm V' = \bm X\bm V\bm X^{\rm T} + \bm Z\bm V_{e}\bm Z^{\rm T}
\end{equation}
For a single-mode loss channel, the symplectic matrix is 
\begin{equation}
    \bm S = \begin{pmatrix}
        \sqrt{\eta}\mathbf{1}_{2} & \sqrt{1-\eta}\mathbf{1}_{2}\\
        -\sqrt{1-\eta}\mathbf{1}_{2} & \sqrt{\eta}\mathbf{1}_{2}
    \end{pmatrix}.
\end{equation}
Thus, $\bm X = \sqrt{\eta}\mathbf{1}_{2}$ and $\bm Z = \sqrt{1-\eta}\mathbf{1}_{2}$, and the final system covariance matrix is $\bm V_{\text{post-loss}} = \eta \bm V_s+(1-\eta)\bm V_{e}$, which is the familiar single-mode result. For the two-subsystem case with loss acting on both channels, we have
\begin{equation}
    \bm S = \begin{pmatrix}
        \sqrt{\eta_{A}}\mathbf{1}_{2} & 0 & \sqrt{1-\eta_{A}}\mathbf{1}_{2} &0\\
        0 & \sqrt{\eta_{B}}\mathbf{1}_{2} & 0 & \sqrt{1-\eta_{B}}\mathbf{1}_{2}\\
        -\sqrt{1-\eta_{A}}\mathbf{1}_{2} & 0 & \sqrt{\eta_{A}}\mathbf{1}_{2} & 0 \\
        0 & -\sqrt{1-\eta_{B}}\mathbf{1}_{2} & 0 & \sqrt{\eta_{B}}\mathbf{1}_{2}
    \end{pmatrix}
\end{equation}
so that
\begin{equation}
\begin{aligned}
    \bm X &= \begin{pmatrix}
        \sqrt{\eta_{A}}\mathbf{1}_{2} & 0 \\
        0 & \sqrt{\eta_{B}}\mathbf{1}_{2}
    \end{pmatrix}\\
    \bm Z &= \begin{pmatrix}
        \sqrt{1-\eta_{A}}\mathbf{1}_{2} & 0 \\
        0 & \sqrt{1-\eta_{B}}\mathbf{1}_{2}
    \end{pmatrix}.
\end{aligned}
\end{equation}
If we write the pre-loss covariance matrix as
\begin{equation}
\label{xpxp 2 mode covariance matrix}
    \bm V_s = \begin{pmatrix}
        \bm V_{s,A} & \bm V_{s,AB}\\
        \bm V_{s,AB}^{\rm T} & \bm V_{s,B}
    \end{pmatrix},
\end{equation}
then the post-loss covariance matrix is
\begin{equation}
    \bm V_{\text{post-loss}} = 
    \begin{pmatrix}
        \eta_{A} \bm V_{s,A} +(1-\eta_{A})\bm V_{e,A} & \sqrt{\eta_{A}\eta_{B}}\bm V_{s,AB}\\
        \sqrt{\eta_{A}\eta_{B}}\bm V_{s,AB}^{T} & \eta_{B}\bm V_{s,B}+(1-\eta_{B})\bm V_{e,B}
    \end{pmatrix}
\end{equation}
Here, $\bm V_{e,j} = (2\bar{n}_j+1)\mathbf{1}_{2}$ for $j\in\{A,B\}$ where $\bar{n}_j$ is the mean number of environmental photons. In practice, one typically takes $\bar n_j=0$, corresponding to a vacuum noise (i.e., a pure-loss channel).

\twocolumngrid

\bibliography{reference_edited}

@article{bose2017,
  title = {Spin Entanglement Witness for Quantum Gravity},
  author = {Bose, Sougato and Mazumdar, Anupam and Morley, Gavin W. and Ulbricht, Hendrik and Toro\ifmmode \check{s}\else \v{s}\fi{}, Marko and Paternostro, Mauro and Geraci, Andrew A. and Barker, Peter F. and Kim, M. S. and Milburn, Gerard},
  journal = {Phys. Rev. Lett.},
  volume = {119},
  issue = {24},
  pages = {240401},
  numpages = {6},
  year = {2017},
  month = {Dec},
  publisher = {American Physical Society},
  doi = {10.1103/PhysRevLett.119.240401},
  url = {https://link.aps.org/doi/10.1103/PhysRevLett.119.240401}
}

@article{marletto2017,
    author = "Marletto, Chiara and Vedral, Vlatko",
    title = "{Gravitationally-induced entanglement between two massive particles is sufficient evidence of quantum effects in gravity}",
    eprint = "1707.06036",
    archivePrefix = "arXiv",
    primaryClass = "quant-ph",
    doi = "10.1103/PhysRevLett.119.240402",
    journal = "Phys. Rev. Lett.",
    volume = "119",
    number = "24",
    pages = "240402",
    year = "2017"
}

@article{marletto2020,
  title = {Witnessing nonclassicality beyond quantum theory},
  author = {Marletto, Chiara and Vedral, Vlatko},
  journal = {Phys. Rev. D},
  volume = {102},
  issue = {8},
  pages = {086012},
  numpages = {8},
  year = {2020},
  month = {Oct},
  publisher = {American Physical Society},
  doi = {10.1103/PhysRevD.102.086012},
  url = {https://link.aps.org/doi/10.1103/PhysRevD.102.086012}
}

@article{marletto2025,
  title = {Quantum-information methods for quantum gravity laboratory-based tests},
  author = {Marletto, Chiara and Vedral, Vlatko},
  journal = {Rev. Mod. Phys.},
  volume = {97},
  issue = {1},
  pages = {015006},
  numpages = {29},
  year = {2025},
  month = {Mar},
  publisher = {American Physical Society},
  doi = {10.1103/RevModPhys.97.015006},
  url = {https://link.aps.org/doi/10.1103/RevModPhys.97.015006}
}

@misc{marletto2025-2,
      title={Classical gravity cannot mediate entanglement},
      author={Marletto, Chiara and Oppenheim, Jonathan and Vedral, Vlatko and Wilson, Elizabeth},
      year={2025},
      eprint={2511.07348},
      archivePrefix={arXiv},
      primaryClass={quant-ph}
}

@article{marshman2020,
  title = {Locality and entanglement in table-top testing of the quantum nature of linearized gravity},
  author = {Marshman, Ryan J. and Mazumdar, Anupam and Bose, Sougato},
  journal = {Phys. Rev. A},
  volume = {101},
  issue = {5},
  pages = {052110},
  numpages = {13},
  year = {2020},
  month = {May},
  publisher = {American Physical Society},
  doi = {10.1103/PhysRevA.101.052110},
  url = {https://link.aps.org/doi/10.1103/PhysRevA.101.052110}
}

@article{schut2024,
  title = {Micrometer-size spatial superpositions for the QGEM protocol via screening and trapping},
  author = {Schut, Martine and Geraci, Andrew and Bose, Sougato and Mazumdar, Anupam},
  journal = {Phys. Rev. Res.},
  volume = {6},
  issue = {1},
  pages = {013199},
  numpages = {12},
  year = {2024},
  month = {Feb},
  publisher = {American Physical Society},
  doi = {10.1103/PhysRevResearch.6.013199},
  url = {https://link.aps.org/doi/10.1103/PhysRevResearch.6.013199}
}

@article{krisnanda2020,
  author  = {Krisnanda, Tanjung and Tham, Guo Yao and Paternostro, Mauro and Paterek, Tomasz},
  title   = {Observable quantum entanglement due to gravity},
  journal = {npj Quantum Information},
  volume  = {6},
  pages   = {12},
  year    = {2020},
  doi     = {10.1038/s41534-020-0243-y},
  url     = {https://doi.org/10.1038/s41534-020-0243-y}
}

@article{qvarfort2020,
  doi = {10.1088/1361-6455/abbe8d},
  url = {https://doi.org/10.1088/1361-6455/abbe8d},
  year = {2020},
  month = {nov},
  publisher = {IOP Publishing},
  volume = {53},
  number = {23},
  pages = {235501},
  author = {Qvarfort, Sofia and Bose, Sougato and Serafini, Alessio},
  title = {Mesoscopic entanglement through central–potential interactions},
  journal = {Journal of Physics B: Atomic, Molecular and Optical Physics}
}

@article{carney2021,
  title = {Using an Atom Interferometer to Infer Gravitational Entanglement Generation},
  author = {Carney, Daniel and M\"uller, Holger and Taylor, Jacob M.},
  journal = {PRX Quantum},
  volume = {2},
  issue = {3},
  pages = {030330},
  numpages = {16},
  year = {2021},
  month = {Aug},
  publisher = {American Physical Society},
  doi = {10.1103/PRXQuantum.2.030330},
  url = {https://link.aps.org/doi/10.1103/PRXQuantum.2.030330}
}

@article{matsumura2022,
  title = {Leggett-Garg inequalities for testing quantumness of gravity},
  author = {Matsumura, Akira and Nambu, Yasusada and Yamamoto, Kazuhiro},
  journal = {Phys. Rev. A},
  volume = {106},
  issue = {1},
  pages = {012214},
  numpages = {11},
  year = {2022},
  month = {Jul},
  publisher = {American Physical Society},
  doi = {10.1103/PhysRevA.106.012214},
  url = {https://link.aps.org/doi/10.1103/PhysRevA.106.012214}
}

@article{balushi2018,
  title = {Optomechanical quantum Cavendish experiment},
  author = {Al Balushi, Abdulrahim and Cong, Wan and Mann, Robert B.},
  journal = {Phys. Rev. A},
  volume = {98},
  issue = {4},
  pages = {043811},
  numpages = {9},
  year = {2018},
  month = {Oct},
  publisher = {American Physical Society},
  doi = {10.1103/PhysRevA.98.043811},
  url = {https://link.aps.org/doi/10.1103/PhysRevA.98.043811}
}

@article{matsumura2020,
  title = {Gravity-induced entanglement in optomechanical systems},
  author = {Matsumura, Akira and Yamamoto, Kazuhiro},
  journal = {Phys. Rev. D},
  volume = {102},
  issue = {10},
  pages = {106021},
  numpages = {12},
  year = {2020},
  month = {Nov},
  publisher = {American Physical Society},
  doi = {10.1103/PhysRevD.102.106021},
  url = {https://link.aps.org/doi/10.1103/PhysRevD.102.106021}
}

@article{miki2022,
  title = {Non-Gaussian entanglement in gravitating masses: The role of cumulants},
  author = {Miki, Daisuke and Matsumura, Akira and Yamamoto, Kazuhiro},
  journal = {Phys. Rev. D},
  volume = {105},
  issue = {2},
  pages = {026011},
  numpages = {13},
  year = {2022},
  month = {Jan},
  publisher = {American Physical Society},
  doi = {10.1103/PhysRevD.105.026011},
  url = {https://link.aps.org/doi/10.1103/PhysRevD.105.026011}
}

@article{miao2020,
  title = {Quantum correlations of light mediated by gravity},
  author = {Miao, Haixing and Martynov, Denis and Yang, Huan and Datta, Animesh},
  journal = {Phys. Rev. A},
  volume = {101},
  issue = {6},
  pages = {063804},
  numpages = {7},
  year = {2020},
  month = {Jun},
  publisher = {American Physical Society},
  doi = {10.1103/PhysRevA.101.063804},
  url = {https://link.aps.org/doi/10.1103/PhysRevA.101.063804}
}

@article{datta2021,
  doi = {10.1088/2058-9565/ac1adf},
  url = {https://doi.org/10.1088/2058-9565/ac1adf},
  year = {2021},
  month = {aug},
  publisher = {IOP Publishing},
  volume = {6},
  number = {4},
  pages = {045014},
  author = {Datta, Animesh and Miao, Haixing},
  title = {Signatures of the quantum nature of gravity in the differential motion of two masses},
  journal = {Quantum Science and Technology}
}

@article{miki2024,
  title = {Quantum signature of gravity in optomechanical systems with conditional measurement},
  author = {Miki, Daisuke and Matsumura, Akira and Yamamoto, Kazuhiro},
  journal = {Phys. Rev. D},
  volume = {109},
  issue = {6},
  pages = {064090},
  numpages = {12},
  year = {2024},
  month = {Mar},
  publisher = {American Physical Society},
  doi = {10.1103/PhysRevD.109.064090},
  url = {https://link.aps.org/doi/10.1103/PhysRevD.109.064090}
}

@article{miki20242,
  title = {Feasible generation of gravity-induced entanglement by using optomechanical systems},
  author = {Miki, Daisuke and Matsumura, Akira and Yamamoto, Kazuhiro},
  journal = {Phys. Rev. D},
  volume = {110},
  issue = {2},
  pages = {024057},
  numpages = {5},
  year = {2024},
  month = {Jul},
  publisher = {American Physical Society},
  doi = {10.1103/PhysRevD.110.024057},
  url = {https://link.aps.org/doi/10.1103/PhysRevD.110.024057}
}

@article{mari2025,
  title = {Can gravity mediate the transmission of quantum information?},
  author = {Mari, Andrea and Zippilli, Stefano and Vitali, David},
  journal = {Phys. Rev. D},
  volume = {113},
  issue = {2},
  pages = {L021905},
  numpages = {8},
  year = {2026},
  month = {Jan},
  publisher = {American Physical Society},
  doi = {10.1103/pfvz-fd54},
  url = {https://link.aps.org/doi/10.1103/pfvz-fd54}
}

@article{tang2025,
  title = {Optimal form factors for experimental proposals on gravity-induced entanglement},
  author = {Tang, Ziqian and Xue, Hanyu and Han, Zizhao and Kan, Zikuan and Li, Zeji and Liu, Yulong},
  journal = {Phys. Rev. D},
  volume = {112},
  issue = {4},
  pages = {042004},
  numpages = {13},
  year = {2025},
  month = {Aug},
  publisher = {American Physical Society},
  doi = {10.1103/jznw-q3q8},
  url = {https://link.aps.org/doi/10.1103/jznw-q3q8}
}

@misc{matsumoto2025,
      title={Space-based cm/kg-scale Laser Interferometer for Quantum Gravity}, 
      author={Nobuyuki Matsumoto and Katsuta Sakai and Kosei Hatakeyama and Kiwamu Izumi and Daisuke Miki and Satoshi Iso and Akira Matsumura and Kazuhiro Yamamoto},
      year={2025},
      eprint={2507.12899},
      archivePrefix={arXiv},
      primaryClass={gr-qc},
      url={https://arxiv.org/abs/2507.12899}, 
}

@article{direkci2025universality,
  title={Universality of stationary entanglement in an optomechanical system driven by non-Markovian noise and squeezed light},
  author={Direkci, Su and Winkler, Klemens and Gut, Corentin and Aspelmeyer, Markus and Chen, Yanbei},
  journal={Physical Review Letters},
  volume={135},
  number={15},
  pages={153601},
  year={2025},
  publisher={APS}
}

@article{direkci2024macroscopic,
  title={Macroscopic quantum entanglement between an optomechanical cavity and a continuous field in presence of non-Markovian noise},
  author={Direkci, Su and Winkler, Klemens and Gut, Corentin and Hammerer, Klemens and Aspelmeyer, Markus and Chen, Yanbei},
  journal={Physical Review Research},
  volume={6},
  number={1},
  pages={013175},
  year={2024},
  publisher={APS}
}

@article{direkci2025characterizing,
  title = {Characterizing stationary optomechanical entanglement in the presence of non-Markovian noise},
  author = {Direkci, Su and Winkler, Klemens and Gut, Corentin and Aspelmeyer, Markus and Chen, Yanbei},
  journal = {Phys. Rev. A},
  volume = {112},
  issue = {4},
  pages = {043512},
  numpages = {20},
  year = {2025},
  month = {Oct},
  publisher = {American Physical Society},
  doi = {10.1103/gysx-m9p2},
  url = {https://link.aps.org/doi/10.1103/gysx-m9p2}
}

@article{yanbei2013,
  doi = {10.1088/0953-4075/46/10/104001},
  url = {https://doi.org/10.1088/0953-4075/46/10/104001},
  year = {2013},
  month = {may},
  publisher = {IOP Publishing},
  volume = {46},
  number = {10},
  pages = {104001},
  author = {Chen, Yanbei},
  title = {Macroscopic quantum mechanics: theory and experimental concepts of optomechanics},
  journal = {Journal of Physics B: Atomic, Molecular and Optical Physics}
}

@article{vanner2011,
  author = {M. R. Vanner  and I. Pikovski  and G. D. Cole  and M. S. Kim and \v{C}. Brukner and K. Hammerer and G. J. Milburn and M. Aspelmeyer},
  title = {Pulsed quantum optomechanics},
  journal = {Proceedings of the National Academy of Sciences},
  volume = {108},
  number = {39},
  pages = {16182-16187},
  year = {2011},
  doi = {10.1073/pnas.1105098108},
  URL = {https://www.pnas.org/doi/abs/10.1073/pnas.1105098108}}

@article{hofer2011,
  title = {Quantum entanglement and teleportation in pulsed cavity optomechanics},
  author = {Hofer, Sebastian G. and Wieczorek, Witlef and Aspelmeyer, Markus and Hammerer, Klemens},
  journal = {Phys. Rev. A},
  volume = {84},
  issue = {5},
  pages = {052327},
  numpages = {10},
  year = {2011},
  month = {Nov},
  publisher = {American Physical Society},
  doi = {10.1103/PhysRevA.84.052327},
  url = {https://link.aps.org/doi/10.1103/PhysRevA.84.052327}
}

@phdthesis{bassam2019,
  title  = {Testing Alternative Theories of Quantum Mechanics with Optomechanics, and Effective Modes for Gaussian Linear Optomechanics},
  author = {Helou, Bassam Mohamad},
  school = {California Institute of Technology},
  year   = {2019},
  doi    = {10.7907/KJ1K-9268},
  url    = {https://resolver.caltech.edu/CaltechTHESIS:12182018-142547647}
}

@article{jordan2024,
  title = {Testing quantum gravity using pulsed optomechanical systems},
  author = {Wilson-Gerow, Jordan and Chen, Yanbei and Stamp, P. C. E.},
  journal = {Phys. Rev. D},
  volume = {109},
  issue = {6},
  pages = {064078},
  numpages = {24},
  year = {2024},
  month = {Mar},
  publisher = {American Physical Society},
  doi = {10.1103/PhysRevD.109.064078},
  url = {https://link.aps.org/doi/10.1103/PhysRevD.109.064078}
}

@article{Brady_2024,
   title={Advances in bosonic quantum error correction with Gottesman–Kitaev–Preskill Codes: Theory, engineering and applications},
   volume={93},
   ISSN={0079-6727},
   url={http://dx.doi.org/10.1016/j.pquantelec.2023.100496},
   DOI={10.1016/j.pquantelec.2023.100496},
   journal={Progress in Quantum Electronics},
   publisher={Elsevier BV},
   author={Brady, Anthony J. and Eickbusch, Alec and Singh, Shraddha and Wu, Jing and Zhuang, Quntao},
   year={2024},
   month=jan, pages={100496} }

@article{horodecki2003,
    author = {Horodecki, Michael and Shor, Peter W. and Ruskai, Mary Beth},
    title = {Entanglement Breaking Channels},
    journal = {Reviews in Mathematical Physics},
    volume = {15},
    number = {06},
    pages = {629-641},
    year = {2003},
    doi = {10.1142/S0129055X03001709}
}

@article{holevo2008,
    author = {Holevo, A. S.},
    title = {Entanglement-breaking channels in infinite dimensions},
    journal = {Problems of Information Transmission},
    volume = {44},
    pages = {171-184},
    year = {2008},
    doi = {10.1134/S0032946008030010}
}

@article{moravcikova2010,
    doi = {10.1088/1751-8113/43/27/275306},
    url = {https://doi.org/10.1088/1751-8113/43/27/275306},
    year = {2010},
    month = {jun},
    volume = {43},
    number = {27},
    pages = {275306},
    author = {Morav\v{c}\'ikov\'a, Lenka and Ziman, Mário},
    title = {Entanglement-annihilating and entanglement-breaking channels},
    journal = {Journal of Physics A: Mathematical and Theoretical}
}

@article{filippov2014,
  title = {Entanglement sensitivity to signal attenuation and amplification},
  author = {Filippov, Sergey N. and Ziman, M\'ario},
  journal = {Phys. Rev. A},
  volume = {90},
  issue = {1},
  pages = {010301},
  numpages = {5},
  year = {2014},
  month = {Jul},
  publisher = {American Physical Society},
  doi = {10.1103/PhysRevA.90.010301},
  url = {https://link.aps.org/doi/10.1103/PhysRevA.90.010301}
}

@misc{kafri2013,
      title={A noise inequality for classical forces},
      author={Kafri, Dvir and Taylor, J. M.},
      year={2013},
      eprint={1311.4558},
      archivePrefix={arXiv},
      primaryClass={quant-ph}
}

@article{kafri2014,
  doi = {10.1088/1367-2630/16/6/065020},
  url = {https://doi.org/10.1088/1367-2630/16/6/065020},
  year = {2014},
  month = {jun},
  publisher = {IOP Publishing},
  volume = {16},
  number = {6},
  pages = {065020},
  author = {Kafri, D and Taylor, J M and Milburn, G J},
  title = {A classical channel model for gravitational decoherence},
  journal = {New Journal of Physics}
}

@misc{alfred2026,
      title={Universal Bound for Entanglement Generation}, 
      author={Alfred Li and Daisuke Miki and Yanbei Chen},
      year={2026},
      eprint={in preparation},
      archivePrefix={arXiv},
      primaryClass={quant-ph}
}

@article{mari2016,
  title={Experiments testing macroscopic quantum superpositions must be slow},
  author={Mari, A. and De Palma G. and Giovannetti, V.},
  journal={Scientific Reports},
  volume={6},
  pages={22777},
  year={2016}
}

@article{belenchia2018,
  title = {Quantum superposition of massive objects and the quantization of gravity},
  author = {Belenchia, Alessio and Wald, Robert M. and Giacomini, Flaminia and Castro-Ruiz, Esteban and Brukner, \v{C}aslav and Aspelmeyer, Markus},
  journal = {Phys. Rev. D},
  volume = {98},
  issue = {12},
  pages = {126009},
  numpages = {9},
  year = {2018},
  month = {Dec},
  publisher = {American Physical Society},
  doi = {10.1103/PhysRevD.98.126009},
  url = {https://link.aps.org/doi/10.1103/PhysRevD.98.126009}
}

@article{danielson2022,
  title = {Gravitationally mediated entanglement: Newtonian field versus gravitons},
  author = {Danielson, Daine L. and Satishchandran, Gautam and Wald, Robert M.},
  journal = {Phys. Rev. D},
  volume = {105},
  issue = {8},
  pages = {086001},
  numpages = {11},
  year = {2022},
  month = {Apr},
  publisher = {American Physical Society},
  doi = {10.1103/PhysRevD.105.086001},
  url = {https://link.aps.org/doi/10.1103/PhysRevD.105.086001}
}

@article{sugiyama2023,
  title = {Quantum uncertainty of gravitational field and entanglement in superposed massive particles},
  author = {Sugiyama, Yuuki and Matsumura, Akira and Yamamoto, Kazuhiro},
  journal = {Phys. Rev. D},
  volume = {108},
  issue = {10},
  pages = {105019},
  numpages = {13},
  year = {2023},
  month = {Nov},
  publisher = {American Physical Society},
  doi = {10.1103/PhysRevD.108.105019},
  url = {https://link.aps.org/doi/10.1103/PhysRevD.108.105019}
}

@article{sugiyama2024,
  title = {Quantumness of the gravitational field: A perspective on monogamy relation},
  author = {Sugiyama, Yuuki and Matsumura, Akira and Yamamoto, Kazuhiro},
  journal = {Phys. Rev. D},
  volume = {110},
  issue = {4},
  pages = {045016},
  numpages = {21},
  year = {2024},
  month = {Aug},
  publisher = {American Physical Society},
  doi = {10.1103/PhysRevD.110.045016},
  url = {https://link.aps.org/doi/10.1103/PhysRevD.110.045016}
}

@article{carney2022,
  title = {Newton, entanglement, and the graviton},
  author = {Carney, Daniel},
  journal = {Phys. Rev. D},
  volume = {105},
  issue = {2},
  pages = {024029},
  numpages = {17},
  year = {2022},
  month = {Jan},
  publisher = {American Physical Society},
  doi = {10.1103/PhysRevD.105.024029},
  url = {https://link.aps.org/doi/10.1103/PhysRevD.105.024029}
}

@article{pedernales2022,
  title = {Enhancing Gravitational Interaction between Quantum Systems by a Massive Mediator},
  author = {Pedernales, Julen S. and Streltsov, Kirill and Plenio, Martin B.},
  journal = {Phys. Rev. Lett.},
  volume = {128},
  issue = {11},
  pages = {110401},
  numpages = {6},
  year = {2022},
  month = {Mar},
  publisher = {American Physical Society},
  doi = {10.1103/PhysRevLett.128.110401},
  url = {https://link.aps.org/doi/10.1103/PhysRevLett.128.110401}
}

@article{kaku2023,
  title = {Enhancement of quantum gravity signal in an optomechanical experiment},
  author = {Kaku, Youka and Fujita, Tomohiro and Matsumura, Akira},
  journal = {Phys. Rev. D},
  volume = {108},
  issue = {10},
  pages = {106014},
  numpages = {18},
  year = {2023},
  month = {Nov},
  publisher = {American Physical Society},
  doi = {10.1103/PhysRevD.108.106014},
  url = {https://link.aps.org/doi/10.1103/PhysRevD.108.106014}
}

@article{fujita2025,
  doi = {10.1088/1361-6382/adf0bb},
  url = {https://doi.org/10.1088/1361-6382/adf0bb},
  year = {2025},
  month = {aug},
  publisher = {IOP Publishing},
  volume = {42},
  number = {16},
  pages = {165003},
  author = {Fujita, Tomohiro and Kaku, Youka and Matsumura, Akira and Michimura, Yuta},
  title = {Inverted oscillators for testing gravity-induced quantum entanglement},
  journal = {Classical and Quantum Gravity}
}

@misc{shiomatsu2025,
      title={Boosting Gravity-Induced Entanglement through Parametric Resonance}, 
      author={Yuka Shiomatsu and Youka Kaku and Akira Matsumura and Tomohiro Fujita},
      year={2025},
      eprint={2511.09169},
      archivePrefix={arXiv},
      primaryClass={gr-qc},
      url={https://arxiv.org/abs/2511.09169}, 
}

@article{hatakeyama2026,
  title = {Theoretical study of the squeezed-light-enhanced sensitivity to gravity-induced entanglement via finite-time analysis},
  author = {Hatakeyama, Kosei and Miki, Daisuke and Yamamoto, Kazuhiro},
  journal = {Phys. Rev. D},
  volume = {113},
  issue = {2},
  pages = {024025},
  numpages = {14},
  year = {2026},
  month = {Jan},
  publisher = {American Physical Society},
  doi = {10.1103/1mfv-y24t},
  url = {https://link.aps.org/doi/10.1103/1mfv-y24t}
}

@article{fukuzumi2026,
  title = {Momentum squeezed state realized via optimal filtering in optomechanics: Implications for gravity-induced entanglement},
  author = {Fukuzumi, Ryotaro and Hatakeyama, Kosei and Miki, Daisuke and Yamamoto, Kazuhiro},
  journal = {Phys. Rev. Res.},
  volume = {8},
  issue = {2},
  pages = {023039},
  numpages = {10},
  year = {2026},
  month = {Apr},
  publisher = {American Physical Society},
  doi = {10.1103/zrs2-sk28},
  url = {https://link.aps.org/doi/10.1103/zrs2-sk28}
}

@article{liu2023,
  title = {Semiclassical gravity phenomenology under the causal-conditional quantum measurement prescription},
  author = {Liu, Yubao and Miao, Haixing and Chen, Yanbei and Ma, Yiqiu},
  journal = {Phys. Rev. D},
  volume = {107},
  issue = {2},
  pages = {024004},
  numpages = {18},
  year = {2023},
  month = {Jan},
  publisher = {American Physical Society},
  doi = {10.1103/PhysRevD.107.024004},
  url = {https://link.aps.org/doi/10.1103/PhysRevD.107.024004}
}

@article{liu2025,
  title = {Semiclassical gravity phenomenology under the causal-conditional quantum measurement prescription. II. Heisenberg picture and apparent optical entanglement},
  author = {Liu, Yubao and Zhong, Wenjie and Chen, Yanbei and Ma, Yiqiu},
  journal = {Phys. Rev. D},
  volume = {111},
  issue = {6},
  pages = {062004},
  numpages = {26},
  year = {2025},
  month = {Mar},
  publisher = {American Physical Society},
  doi = {10.1103/PhysRevD.111.062004},
  url = {https://link.aps.org/doi/10.1103/PhysRevD.111.062004}
}

@article{miki2025,
  title = {Role of quantum measurements when testing the quantum nature of gravity},
  author = {Miki, Daisuke and Kaku, Youka and Liu, Yubao and Ma, Yiqiu and Chen, Yanbei},
  journal = {Phys. Rev. D},
  volume = {111},
  issue = {10},
  pages = {104084},
  numpages = {32},
  year = {2025},
  month = {May},
  publisher = {American Physical Society},
  doi = {10.1103/PhysRevD.111.104084},
  url = {https://link.aps.org/doi/10.1103/PhysRevD.111.104084}
}

@article{zhong2025,
  title = {Distinguishing quantum and classical gravity via nonstationary test mass dynamics},
  author = {Zhong, Wenjie and Liu, Yubao and Ma, Yiqiu},
  journal = {Phys. Rev. D},
  volume = {112},
  issue = {4},
  pages = {044060},
  numpages = {21},
  year = {2025},
  month = {Aug},
  publisher = {American Physical Society},
  doi = {10.1103/nl32-g2r4},
  url = {https://link.aps.org/doi/10.1103/nl32-g2r4}
}

@article{duan2000,
  title = {Inseparability Criterion for Continuous Variable Systems},
  author = {Duan, Lu-Ming and Giedke, G. and Cirac, J. I. and Zoller, P.},
  journal = {Phys. Rev. Lett.},
  volume = {84},
  issue = {12},
  pages = {2722--2725},
  numpages = {0},
  year = {2000},
  month = {Mar},
  publisher = {American Physical Society},
  doi = {10.1103/PhysRevLett.84.2722},
  url = {https://link.aps.org/doi/10.1103/PhysRevLett.84.2722}
}

@article{simon2000,
  title = {Peres-Horodecki Separability Criterion for Continuous Variable Systems},
  author = {Simon, R.},
  journal = {Phys. Rev. Lett.},
  volume = {84},
  issue = {12},
  pages = {2726--2729},
  numpages = {0},
  year = {2000},
  month = {Mar},
  publisher = {American Physical Society},
  doi = {10.1103/PhysRevLett.84.2726},
  url = {https://link.aps.org/doi/10.1103/PhysRevLett.84.2726}
}

@article{palomaki2013,
  author = {T. A. Palomaki  and J. D. Teufel  and R. W. Simmonds  and K. W. Lehnert },
  title = {Entangling Mechanical Motion with Microwave Fields},
  journal = {Science},
  volume = {342},
  number = {6159},
  pages = {710-713},
  year = {2013},
  doi = {10.1126/science.1244563},
  url = {https://www.science.org/doi/abs/10.1126/science.1244563},
}

@article{junxin2020,
  author  = {Chen, J. and Rossi, M. and Mason, D. and Schliesser, A.},
  title   = {Entanglement of propagating optical modes via a mechanical interface},
  journal = {Nature Communications},
  volume  = {11},
  pages   = {943},
  year    = {2020},
  doi     = {10.1038/s41467-020-14768-1}
}

@article{Lami2024,
  title = {Testing the Quantumness of Gravity without Entanglement},
  author = {Lami, Ludovico and Pedernales, Julen S. and Plenio, Martin B.},
  journal = {Phys. Rev. X},
  volume = {14},
  issue = {2},
  pages = {021022},
  numpages = {47},
  year = {2024},
  month = {May},
  publisher = {American Physical Society},
  doi = {10.1103/PhysRevX.14.021022},
  url = {https://link.aps.org/doi/10.1103/PhysRevX.14.021022}
}
\end{document}